# Sensitivity, uncertainty analyses and algorithm selection for Sea Ice Thickness retrieval from Radar Altimeter

Vera Djepa

*DAMTP, University of Cambridge, Cambridge, United Kingdom,*

**ABSTRACT**

Sea Ice Thickness (SIT) is important climate variable, sensitive to environmental changes in the Polar regions. For accurate forecast of climate change, sea ice mass balance, ocean circulation and sea- atmosphere interactions is required to have long term records of SIT.
Different approaches have been applied to retrieve SIT, using ground based, Upward Looking Sonar (ULS) (moored or on Submarine), airborne (Laser altimeters (LA) on board Operational Ice Bridge (OIB)), or satellite observations, where only satellite altimetry, radar or laser, have been proven to provide hemispheric estimates of SIT distribution over a sufficient thickness range. SIT is retrieved from freeboard (F), measured from Radar Altimeter (RA), using equation for hydrostatic equilibrium, where sea ice density and snow depth and density are the main input variables, impacting the accuracy of the derived SIT and Sea Ice Draft (SID). Snow depth and density from Warren Climatology (WC) have been used successfully until now for SIT retrieval from RA (on board of ESA/ERS1, 2, ENVISAT and CryoSat-2) and laser altimeter on board NASA/ICESat (Ice, Cloud and land Elevation Satellite). To simplify the algorithm for SIT retrieval from RA, constant ice density has been applied until now, which lead to different results for derived SIT and SID, in dependence on input information for sea ice density and snow depth.
The purpose of this paper is to select algorithm for SID and SIT retrieval from RA, using statistical, sensitivity analyses and independent observations of SID from moored ULS, or on Submarine. Considering the wide range of ice densities (used for SIT retrieval) and SIT dependence on free board, ice type and snow depth (leading to different SIT and SID estimates from RA), the impact of ice density and snow depth on accuracy of the retrieved SIT has been examined, applying sensitivity analyses, and the propagated uncertainties have been summarised. Methods, uncertainties and accuracy of algorithms for snow depth retrieval in the Arctic have been discussed and it is concluded that the assumption of half snow depth over First Year Ice (FYI) will lead always to underestimation of SIT and SID derived from RA and is not applicable for SIT retrieval from RA, using the equation for hydrostatic equilibrium. Based on statistical, sensitivity analyses and comparison of retrieved SID from RA with collocated SID from ULS, it is concluded that WC is applicable over FYI and Multi Year Ice (MYI), using the equation for hydrostatic equilibrium, integrated with freeboard depended (FD) algorithm for SIT retrieval from RA. Algorithms for freeboard depended ice densities and SIT retrieval from RA have been developed and compared with existing and new algorithms and a FD algorithm for SIT retrieval from RA have been selected, based on statistical, sensitivity analyses and comparison with collocated observations of SID from moored ULS and on Submarine .
ESA (ERS1, 2, ENVISAT, CryoSat2), future ESA (Sentinel) and NASA satellite and airborne missions, climate and numerical forecast programs will benefit the results of this paper.

## Introduction

Sea Ice Thickness (SIT) is an important climate variable, sensitive to environmental changes in the Polar Regions. Sea ice volume, the product of SIT and sea ice area, seems to be an even better indicator of changes in the sea ice cover than SIT or sea ice extent alone. For accurate forecasting of climate change, sea ice mass balance, sea level ocean circulation and ocean-atmosphere interactions it is required to have long-term records of SIT with information about retrieval uncertainties and potential biases.
Ground based (drilling), ULS moored or on submarines [Wadhams, 1992, Drucker et al, 2003] and airborne sensors [Kurtz et al, 2012], can provide information about SIT either directly or via Sea Ice Draft (SID) or freeboard (F) measurements, but the data are on a small scale and not regular. Only satellite observations can provide regular information about SIT and F distribution at the spatial scales required for climate impact study and validation of climate (CM) and weather prediction (WPM) models.

Different approaches have been applied to retrieve SIT from satellite observations, using radar altimeter (RA) aboard ESA (ERS1/2, Envisat and CryoSat) satellites [Connor et al, 2009, Laxon et al, 2012], laser altimeter (GLAS) aboard the ICESat satellite [Kwok and Cunningham, 2008], Synthetic Aperture Radar (SAR) [Kwok, 1995] for thin ice, microwave radiometry (MIRAS) aboard the Soil Moisture and Ocean Salinity (SMOS) satellite for thickness up to 50 cm [Kaleschke, L., et al., 2012], or energy balance models coupled with satellite thermal infrared imagery [Wang, 2010]. From all the different satellite approaches to retrieve SIT only satellite altimetry, radar or laser, have been proven to provide hemispheric estimates of the SIT distribution over a sufficient thickness range [Kwok and Conningham, 2008, ENVISAT, 2010]. Because RA data have also been available for two decades until now, this data are a perfect candidate to derive a long-term SIT distribution data set, which can be continued with the current CryoSat-2 and the planned Sentinel 3 mission.

Sea Ice Thickness (SIT) have been retrieved in the Arctic since 1992 from Satellite Radar Altimeter (RA) on board of ERS1, 2 and ENVISAT assuming hydrostatic equilibrium [Kwok, 2010, Connor et al 2009], or using empirical algorithms [Alexandrov, et al, 2011, Wadhams, 2000]. Conversion of F, measured from RA to SIT and SID is based on empirical relationships, or using equation for hydrostatic equilibrium, where sea ice density and snow depth and density are the main input variables, impacting the accuracy of the derived SIT and SID. To simplify the algorithm for SIT retrieval from RA (assuming hydrostatic equilibrium), fixed ice density and snow depth and density from Warren Climatology (WC) (Warren, 1999) have been used to retrieve SIT from CryoSat2 [Laxon et al, 2012], ERS 1, 2/RA and ENVISAT/RA2 [Connor et al 2009] and ICESat [Kwok and Conningham, 2008], which lead to un-comparative SIT and SID results, in dependence on input sea ice density. The SIT, retrieved from laser altimeter depends on the same input variables as the SIT, retrieved from RA [Kwok, 2010], which has to be considered if SIT from laser and radar altimeter are compared, using different input variables. Using sensitivity analyses and algorithm comparison to retrieve SIT from radar altimeter, the propagated uncertainties have been summarised and it was shown that comparison of SIT, derived from laser and RA, using different input variables of snow depth, density, ice and water density may lead to misleading results.

Laxon et al (2012) applied the equation for hydrostatic equilibrium to retrieve SIT from CryoSat-2 assuming constant ice densities over First Year Ice (FYI) and Multiyear Ice (MYI) and half snow depth from WC over FYI, relaying on limited few tracks of Operational Ice Bridge (OIB) radar observations from Kurtz and Farrell, (2011). The assumption of half snow depth over FYI based on few tracks from OIB radar, or snow depth retrieved from AMSR-E with not proved accuracy [Cavalieri et al, 2012] and confirmed 2.3 times underestimation of the AMSR-E algorithm for snow depth [Worby et al, 2008], have been examined. Using sensitivity, uncertainty analyses and comparison of retrieved SID from RA with collocated SID from ULS observations it was confirmed that the assumption for half snow depth over FYI will lead to systematic underestimation of SIT, retrieved from RA (aboard CryoSat2, ERS1, 2, and ENVISAT). It was demonstrated that the snow depth and density from WC is applicable over FYI and MYI, using the equation for hydrostatic equilibrium and freeboard dependent algorithm for sea ice thickness retrieval from RA.

Considering the ice density dependence on freeboard, ice type and snow depth, two algorithms for freeboard dependent ice density and SIT retrieval from RA have been developed, validated with observations [Ackley et al, 1976, Kovacs, 1996] and sensitivity analyses and compared with existing algorithms and collocated SID from ULS observations. The retrieved SID from RA, using selected algorithms is calculated and compared with SID derived from collocated ULS data in different seasons and locations in the Arctic and the FD algorithm satisfying long term minimum bias and RMS error in different locations and seasons has been selected for F to SIT and SID conversion, assuming hydrostatic equilibrium. Algorithm selection is based on uncertainty, sensitivity analyses and comparison of SID, calculated from RA, applying different algorithms, with collocated data from ULS.

The paper is organised as follows: Section 1 is focused on SID and SIT retrieval from alternative observations (from laser altimeter and ULS) and corresponding uncertainties. Uncertainties, sensitivity analysis of the retrieved SIT from RA and the algorithms are provided in Section 2. Results for algorithm comparison are listed in Section 3. The final algorithm selection, accuracy and conclusions are provided in Section 4.

## 1. Retrieval of SIT and SID from laser altimeter and ULS

This section is focussed on uncertainties of SID and SIT retrieval from ULS on Submarine and laser altimeter on board of OIB. The algorithm and uncertainties to retrieve SIT from laser altimeter is discussed because the retrieved SIT from laser altimeter depends on the same input variables as the SIT, retrieved from RA, which should be considered if SIT from the two instruments is compared.

Independent collocated observations of SID from moored ULS or on submarine have been used for algorithm selection of SIT retrieval from RA, which requires uncertainties and accuracy analyses of the retrieved SID from ULS.

## 1.1. Retrieval of SID from ULS and corresponding uncertainties

Collocated SID, retrieved from moored ULS and on submarine have been used for algorithm selection and comparison with the retrieved SID from RA. The SID (d), measured by sonar transducer, mounted on the submarine, or d from mooring ULS, is calculated from the difference between the depth of the transducer $D_T$ below the sea level and the sonar measured range, r, to the ice bottom by:

$$d = D_T - r ,  \qquad (1)$$

where $D_T=D-H$, H is the vertical distance from the pressure sensor to the sonar transducer H (H=15.7m for US submarines), with the keel depth D, determined by the measured pressure, p, at the submarine [Rothrock and Wensnahan, 2007]. The range, r, is the distance from the ULS on submarine to the ice underside, measured as r=2tc, where 2t is the return travel time of the emitted from ULS acoustic pulse, which is reflected at the ice underside and c is the mean sound speed in the water column. The open water offset (thin ice correction) and the impact of the beam width are the most important biases contribute to the accuracy of the SID, derived from ULS on submarine [Rothrock and Wensnahan, 2007]. A mean bias of 29cm, accounting for the footprint error of ULS with 2$^o$ beam width and open water correction, with standard deviation of 25cm has been reported for NSIDC ULS draft data [Rothrock and Wensnahan, 2007], which is valid for SID from NSIDC in Beaufort Sea from 1996 and a mean bias 0.227m, with standard deviation $\sigma(\varepsilon)$= 0.1824m have been estimated for SID data derived in Beaufort Sea on 2007 (Djepa and Wadhams, 2013).

Collocated SID from the moored ULS, measured in the Beaufort Gyre (BG) in winters 2003-2008, with RA freeboard have been also used for algorithm selection. The BG Exploration Project (BGEP) in 2003, as part of Arctic Observing Network (AON), was organised to test the origin of the salinity minimum in the centre of the BG from bottom ULS moorings and shipboard measurements. The method for SID retrieval from moored ULS is similar to that of ULS on Submarine [Drucker and Seelye, 2003] and the accuracy of the derived SID is 0.05m, up to 0.1m [Krishfield R., A. Proshutinsky, 2006].

The uncertainties of the collocated SID, derived from ULS (in Beaufort Sea, 1996, NSIDC, and in Beaufort Gyre in 2003-2008), used to compare and select the SIT algorithm for conversion of F from RA to SIT, are summarized in Table 1.

Table 1. Uncertainties of the derived SID from collocated ULS, used for algorithm comparison

| Uncertainty | NSIDC (1996) | BGEP(2002-2008) |
|---|---|---|
| $\sigma$(m) | 0.25m | 0.05 |

One can see that ULS provide independent observations of SID with known uncertainties (up to 5cm for BGEP experiment), which makes these data suitable for algorithm comparison and selection, using collocated freeboard from RA. Apart of SID from ULS, SIT from Operational Ice Bridge (OIB), developed to bridge the gap between NASA's Ice, Cloud and Land Elevation Satellite (ICESat) mission and the upcoming ICESat-2 mission are available [http://nsidc.org/data/icebridge/index.html] and the uncertainties of the retrieved SIT from laser altimeter on board of OIB will be examined in the next section.

## 1.2. SIT retrieval from laser altimeter on board OIB and corresponding uncertainties

The SIT, retrieved from the airborne laser altimeter (ATM) on board OIB is snow depth, density, ice density and freeboard dependent as the SIT retrieved from RA [Kurtz et al 2012]. Local sea surface height, estimated from the elevation of nearby leads, is usually subtracted from the ice floe elevation measurements to derive sea ice freeboard. An absence of leads in the ice survey region prevents direct measurement of local sea ice freeboard and the first approximation, not corrected for instantaneous sea surface conditions (e.g., tides, currents, and atmospheric pressure), is obtained by subtracting EGM2008 geoid model from the ice floe elevations, which may lead to uncertainties in freeboard and SIT retrieval from laser altimeter [Kurtz et al, 2012]. Sea ice thickness, $h_i$, is calculated from the freeboard of the airborne laser scanner ($h_f$) at the air-snow interface by:

$$h_i = \rho_w h_f (\rho_w - \rho_i) - (\rho_w - \rho_s) h_s / (\rho_w - \rho_i) \tag{2}$$

where the snow depth ($h_s$), is retrieved from OIB/radar, constant sea ice ($\rho_i$=914.3 kg/m$^3$), water ($\rho_w$=1023.9kg/m$^3$) and snow ($\rho_s$=264.3kg/$^3$) density are used to calculate SIT from ATM in 2009 [Kurtz et al, 2012]. Use of constant different snow densities (264kg/m$^3$ in 2009 expedition and 320kg/m$^3$ during 2010) will impact the accuracy of the derived snow depth from OIB/radar, which is input variable in Equation (2), as well as the accuracy of the retrieved SIT from Laser altimeter. Using constant snow density, not accounting for snow grain and roughness, and considering the strong dependence of the algorithm on snow-ice and snow–air freeboards, may lead to inaccurate snow depth retrieval in locations with different snow density, not precise estimation of $h_{fi}$ or $h_{fs}$ and SIT because the accuracy of SIT, estimated from ATM depends on accuracy of the input variables (ice ($\rho_i$), water ($\rho_w$) and snow ($\rho_s$) density and snow depth) and freeboard. According to Cavalieri et al, (2012) the OIB radar snow depth estimate is calibrated with the AMSR-E snow depth product, but the accuracy of the retrieved snow depth from the both instruments is not known and there is no way to distinguish between first-year ice with a deep snow cover and multiyear ice, which makes questionable the accuracy of the retrieved SIT, applying Equation (2) when snow depth from OIB radar is used. The accuracy of the freeboard impacts also the accuracy of the retrieved $h_i$ by Equation 2. According to Kurtz et al, (2012) an empirically derived offset (0.307 m) is applied as a sea surface height correction to the entire ATM surface elevation data set to obtain corrected sea ice freeboard, which may introduce errors in estimated sea ice freeboard and $h_i$ in areas with different tides, currents or atmospheric pressure than these in the validation points. As the ice density depends on ice type, use of constant ice density (914.3 kg/m$^3$) may lead also to errors in estimated SIT over MYI or FYI under melting conditions, which has been analysed, using sensitivity analyses and comparison with SID retrieved from ULS data [Djepa, 2013]. Considering the unknown accuracy of the input variables in Equation 2, any comparison of SIT, retrieved from laser and radar altimeter, using different input variables ($\rho_i$, $\rho_w$, $\rho_s$, $h_s$ and $h_f$) with unknown accuracy may lead to misleading results.

Considering above, different spatial resolution of freeboard from OIB laser altimeter and RA, the unknown accuracy of the snow depth retrieved from OIB/radar [Cavalieri et al, 2012], the dependence of the SIT retrieved from OIB/laser altimeter on the same variables as the SIT retrieved from RA ($\rho_i$, $\rho_w$, $\rho_s$, $h_s$ and $h_f$), it was concluded that only the ULS collocated observations of SID with known accuracy will be used in this study for algorithm selection to retrieve SIT from RA.

## 2. Retrieval of Sea Ice Thickness from Radar Altimeter

The satellite radar altimeter measures the freeboard of the ice ($h_{fi}$) by subtracting the elevation of the Sea Surface Height (SSH) of the water ($h_{ssh}$) from the observed elevation ($h_{obs}$) [Kwok, 2010]:

$$h_{fi} = h_{obs} - h_{ssH} \tag{3}$$

where the SSH is a sum of contributions from a number of physical processes [ENVISAT, 2010]:

$$h_{ssH}(x, t) = h_g(x) + h_a(x, t) + h_T(x, t) + h_d(x, t) \tag{4}$$

where $h_g$ is associated with geoid undulations, $h_a$ represents the atmospheric pressure loading, $h_T$ is the tidal contributions, and $h_d$ accounts for dynamic topography associated with geostrophic surface currents. All these terms vary in time and space and contribute to the uncertainty of the derived SIT when the sea ice, water and snow height are measured relatively to the level of a reference ellipsoid.

The penetration depth of radar signal depends on snow properties. Beaven et al. (1995) concludes from laboratory experiments, that a Ku-band radar signal at normal incidence reflects at the snow-ice interface if sea ice is covered by dry, cold snow. The radar signal does not penetrate into the snow layer, but reflects from the air-snow interface (Hallikainen, 1992) in presence of wet snow. Internal ice layers and ice lenses in the snow layer, snow grain size and the presence of frost flowers affect also the penetration depth, but it is assumed that the Ku-band radar reflects at the snow-ice interface in presence of cold and dry conditions. Assuming hydrostatic equilibrium and that radar returns are from the snow–ice interface, the SIT, ($h_i$), derived from RA is approximated as a function of snow depth ($h_s$), density of water, $\rho_w$, ice, $\rho_i$, and snow, $\rho_s$, [Kwok, 2010]:

$$h_i = (h_s\rho_s + h_{fi}\rho_w)/(\rho_w - \rho_i), \tag{5}$$

Considering that ice draft can be retrieved from the difference of SIT and F and accounting for Equation 5, we receive the following relationship of SID, $h_i$, $h_s$, $\rho_s$, $\rho_i$, $\rho_w$ and $h_{fi}$:

$$d_{ra} = h_i - h_{fi} = (h_s\rho_s + h_{fi}\rho_i)/(\rho_w - \rho_i), \tag{6}$$

where $h_i$ is the SIT, calculated by Equation (5), assuming that the radar return is on snow-ice interface, $h_f = h_{fi}$. One can see that SID, derived from RA depends on the same input variables and their uncertainties as the SIT derived from laser altimeter (snow depth and density, ice, water density and freeboard).

Equation (5) have been widely applied to retrieve SIT from the freeboard, measured from satellite radar altimetry on board ERS/1, 2, Envisat and CryoSat [Kwok, 2010, Connor et al, 2009, Laxon et al, 2012], assuming constant values for sea ice and water density, snow depth and density taken from Warren climatology [Warren, 1999]. The impact of the uncertainties of the input variables ($h_s$, $\rho_s$, $\rho_i$, $\rho_w$ and $h_{fi}$) will be discussed in the next section.

## 2.1. Impact of the input variables on accuracy of the derived SIT from RA

Snow depth and density and water and ice densities are the main contributing factors to the accuracy of the retrieved SIT and SID from RA.

### 2.1.1. **Snow depth and density and corresponding uncertainties in the Arctic**

Snow depth and density from WC [Warren, 1999] have been used from many authors [Laxon, 1912, Kwok and Cunningham, 2008] to retrieve SIT from laser or radar altimeter.

**Snow depth and density from Warren Climatology (WC)**

The snow depth and ice density, estimated from WC have been used successfully until now for SIT retrieval from RA on board ERS1, ERS2, ENVISAT and ICESat [Laxon, et al, 2012, Kwok and Conningham, 2008], using the equations of hydrostatic equilibrium over FYI and MYI in the Arctic. The snow depth $h_s$ (cm) from WC, is calculated by two-dimensional quadratic fit to mean monthly measured snow depth, $H_0$, at the North Pole:

$$h_s = H_0 + Ax + By + CxyDx^2 + Ey^2, \tag{7}$$

where x (latitude) and y (longitude) are positive axis respectively along 0°N and 90°E in degrees and the coefficients A, B, C, D, E, the monthly dependence of RMS error (ε) of the fit (in cm), the slope F of the trend lines in cm yr$^{-1}$, inter-annual variability (IAV) and mean snow depth uncertainty, $\sigma_{hs} = 0.075$ cm (for winter months) are given in [Warren et al., 1999].

The snow density in WC varies seasonally and as the seasons change from autumn to winter the snow density increases from ~250 kg m$^{-3}$ in September to ~ 320 kg m$^{-3}$ in May, due to the effects of the snow settling and wind, with the highest snow density during snow melt and mean $\rho_s$=295 kg/m$^3$, $\sigma(\rho_s)$= 4.4kg/m$^3$, during the winter months [Radionov et al, 1997]. Alexandrov et al. (2011) examine snow density from the Sever expeditions and found an average snow density on FYI of 324 ± 50 kg m$^{-3}$, which is similar to $\rho_s$ = 310-320 kg m$^{-3}$, estimated by WC over MYI. This confirms that there are not differences in the snow depth over FYI and MYI considering that the snow water equivalent (SWE) is a snow depth multiplied by snow density and equal snow densities over FYI and MYI will lead to similar snow depth over FYI and MYI for the same SWE. WC have been validated with in-situ observations from different authors [e.g. Radionov et al, 1997] in different locations in the Arctic and have been successfully applied for SIT retrieval from RA on board ERS1, ERS2, ENVISAT, CryoSat2 and ICESat [Connor, et al, 2009, Kwok and Cunningham, 2008, Laxon et al 2012]. Considering above, collocated snow depth and density from WC with the same spatial resolution as the RA averaged area have been used for SIT algorithm validation and selection.

**Uncertainties in the Snow depth in the Arctic, retrieved from airborne and satellite observations**

Higher spatial resolution snow depth estimates with low or unknown accuracy and many restrictions for application have been available recently from AMSR-E and OIB radar. Unfortunately the empirical AMSR-E algorithm for snow depth retrieval is with number of limitations [Cavalieri et al, 2012] and validation with surface observations demonstrates about 2.3 times underestimation of the snow depth, retrieved from AMSR-E microwave observations [Worby, et al, 2008].

**Limitations and uncertainties of the snow depth retrieved from AMSR-E**

The algorithm for retrieving snow depth on sea ice from AMSR-E satellite passive microwave data is based on an empirical relationship between in situ snow depths and the ratio of the normalised difference of brightness temperatures, measured by AMSRE at 37GHz and 19GHz), assuming that the scattering increases if the snow depth increases and that the scattering efficiency is greater at 37 GHz than at 19 GHz, leading to increase of the difference between these frequencies when the snow depth increases [Comiso, 1997]:

hs = *a* + *b* GR                                                                (8)

where GR is the gradient ratio of vertically-polarized brightness temperatures (TB), compensated for sea ice concentration:

GR =(T37VB − T19VB)/(T37VB + T19VB )                                             (9)

The set of (*a; b*) coefficients are derived from brightness temperatures measured by SSM/I and snow depths, collected on Antarctic smooth FYI with a correlation coefficient of -0.77. These coefficients are calibrated for AMSR-E (2.9; -782), and are applied in both hemispheres [Markus and Cavalieri, 2004], which may introduce errors in the derived snow depth in dependence on local conditions. The AMSR-E snow depth algorithm is with number of limitations [Markus and Cavalieri, 2004, Cavalieri et al, 2012] and the accuracy of the retrieved snow depth may decrease due to:
- Use of empirical coefficients, derived for SSMI in Antarctica;
- Missing brightness temperatures;
- Use of constant open water points from SSM/I algorithm for AMSR-E snow depth algorithm;
- Errors due to atmospheric impact or estimation of the Sea Ice Concentration (SIC).

The AMSR-E snow depth algorithm is not applicable for SIC<0.2m, in presence of snow melting areas and it does not consider the impact of ice roughness. Uncertainties due to: i) errors from snow metamorphism; ii) changes in atmospheric water vapour; iii) present of land; iv) presence of MYI and changes in snow density impact the accuracy of the retrieved snow depth from AMSR-E. The AMSR-E snow depth product is applicable for snow-depth up to 0.45m only and the above restrictions lead to underestimation of the snow depth [Worby, et al, 2008], retrieved from AMSR-E, which increase as the sea ice concentration decreases and may lead to misleading conclusions for decreased snow depth over FYI. Worby, et al, (2008) estimated 2.3 times underestimation of the snow depth, retrieved from AMSR-E in comparison with surface observations. Considering above restrictions and different spatial resolution of AMSR-E and RA, the snow depth retrieved from AMSR-E cannot be used for quantitative assessment of the snow depth over FYI and is not applicable for SIT retrieval from RA.

**Snow depth from Operation Ice Bridge (OIB), uncertainties and limitations.**

The OIB snow radar has 14.5 m × 11 m spatial resolution and after averaging of ~40 radar measurements the OIB snow depth product is with resolution 40mx11m with expected snow depth retrievals in the range 0.05 – 1.2 m [Kurtz et al., 2012]. Snow depth retrievals from the OIB snow radar depend on accuracy of detection of air–snow and snow–ice interfaces and the distance between the two interfaces, which depends on dielectric constant of snow density. Very low constant snow density (264kg/m$^3$) has been used from Farrell, et al, (2012) to retrieve the snow depth in April 2009, considering that the mean snow density from WC for this month is $\rho_s$ =320kg/m$^3$ [Warren et al,1999] and the mean snow density for FYI is 324kg/m$^3$ [Alexandrov, et al, 2010] and vary up to 430kg/m$^3$, which impacts the accuracy of the retrieved snow depth. Due to the relatively low difference between the dielectric constants for air and snow, as well as surface roughness effects, the air–snow interface is difficult to detect with OIB/radar and a threshold is set to identify the top of the snow layer within the radar return, which depends on ice snow interface and measured standard deviation, leading to snow depth dependence on sea ice freeboard [Kurtz et al, 2012]. The snow depth from the OIB radar depends on the distance between top snow air interface and bottom ice snow interface and speed of light, which again depends on snow density. Using not correct, constant $\rho_s$, not accounting for snow grain and roughness, and considering the strong dependence of the algorithm on snow -ice and snow –air freeboards, may lead to inaccurate snow depth retrieval in locations with different snow density or not precise estimation of $h_{fi}$ or $h_{fs}$. Due to calibration of the OIB radar with the AMSR-E airborne simulator, Cavalieri et al, (2012) concluded that it is not possible to provide accuracy of the snow depth retrieved from the airborne radar and AMSR-E because the accuracy of any of these algorithms is not known. Cavaliery et al [2012] concluded also that despite the airborne AMSR-E and OIB/radar are calibrated to show similar readings, comparison of the snow depth derived from AMSR-E satellite observations with $h_s$(OIB) may lead to higher biases in some locations, not only due to atmospheric and spatial resolution impact but also due to algorithm problems. According to Cavalieri, et al(2012) and Kurtz, et al (2012), the snow depth estimates from OIB/radar and AMSR-E cannot be used for quantitative assessment of the snow depth over FYI because it is ongoing validation and improvement of the OIB/radar and AMSRE-E snow depth products and some validation studies [Worby, et al, 2008] demonstrates 2.3 times underestimation of the snow depth, retrieved from AMSR-E compared with the collocated surface observations.
Considering above limitations, different spatial resolutions, availability of OIB and AMSR-E snow depth products in limited locations and time, the only long term source for snow depth and snow density estimation in the Arctic, collocated with RA observations and validated with surface data with available monthly uncertainties is the WC, which has been used until now to calculate SIT from RA and will be used in this study for algorithm comparison.

2.1.2. **Sea Ice and water density and uncertainties**

Sea water density, $\rho_w$ , depends on salinity, S, temperature, T, and pressure: $\rho_\omega = \rho_\omega$ (S, T, p) (kg/m$^3$) and ranges from about 1022 kg/m$^3$ at the sea surface to 1050 kg/m$^3$ at the bottom of the ocean, mainly due to compression. The density of sea water across the Beaufort Shelf and slope off Alaska varies between 1023.2 kg/m$^3$ in October to 1024.2 kg/m$^3$ in April and a mean value of $\rho_w$ =1024kg/m$^3$ has been used for SIT retrieval from ICESat [Kwok and Cunningham, 2008]. Water density of $\rho\omega$ =1030kg/m$^3$ has been used to calculate SIT from RA and CryoSat2 [Laxon et al, 2012]. Water density $\rho\omega$ =1025+0.5kg/m$^3$ has been used to calculate mean MYI sea ice density (882 kg/m$^3$) [Alexandrov et al, 2010].

The density of sea ice depends on density of pure ice, the fractional volume of air pockets and the amount and density of brine in the ice. The density of pure ice at 0°C is 916.4kg m$^{-3}$ [Hobbs, 1974] and is increasing to 919.3 kg/m$^3$ at -30°C. The density of sea ice can be greater than these values because of the effect of brine inclusions in the ice, or less because of the effect of air bubbles. The brine volume generally increases with temperature, impacting the ice density, which depends on ice type. Timco and Frederking (1996) reported that FY ice density is typically between 840 and 910 kgm$^{-3}$, while MY ice density is between 720 and 910 kgm$^{-3}$ and according to Schwarz and Weeks (1977) the typical FYI density is in the range 910-920kg m$^{-3}$. Vinjea and Finnekasa (1986) drilled 382 holes in level ice of different ages in Fram Strait during July-August and obtained a mean ice density of 902 kg/m$^3$, which was estimated in summer period with impact of melting. According to Wadhams, et al, (1992) FY ice density range is 910-920kg/m$^3$, but it depends on temperature, free-board, snow depth, presence of melting. Densities of MY and FY ice samples taken below the waterline are not significantly different, and both ice types have typical values between 920 and 940 kgm$^{-3}$ [Wadhams, et al, 1992, Alexandrov et al, 2010]. This difference is mainly due to the higher volume of air-filled pores in MYI compared to FYI.

The sea ice density dependence on SIT, $h_{fi}$, $h_s$ and $\rho_s$ has been documented from many authors [Wadhams et al, 1992, Kovacs, 1996] and using the equation for isostatic equilibrium, sea ice density has been calculated from different authors as a function of $h_{fi}$, $h_s$ and $\rho_s$ [Ackley, 1976, Alexandrov et al, 2010]. Assuming that the ice is in isostatic equilibrium (Equation (5)), snow density ($\rho_s$ =324 $\pm$50kgm$^{-3}$), mean snow depth ($h_s$) (0.05m), ice thickness ($h_i$) and ice freeboard ($h_{fi}$) from Sever measurements, [Alexandrov et al, 2010] estimated mean ice density for FY ice $\rho_i$ =916.7±35.7 kgm$^{-3}$ from:

$$\rho_i = \rho_w - (h_{fi}\rho_w + \rho_s h_s)/h_i, \qquad (10)$$

Where the water density is $\rho_w$ =1025$\pm$0.5kg/m$^3$. Because ice density depends on $h_{fi}$, $h_s$ and $\rho_s$, the FY ice density in different locations and snow depth may be different than $\rho_i$ =916.7kg/m$^3$, calculated from observations during Sever expedition.

By inserting density values for the upper and lower ice layers, using freeboard (0.3 m) and ice thickness 2.9 m, a mean MY ice density 882±23 kg/m$^3$ is calculated by Alexandrov et al, 2010 from:

$$\rho_i = \rho_{il}(1 - h_{fi}/h_i) + \rho_{iu} h_{fi}/h_i \qquad (11)$$

The estimated MYI density 882 kg/m$^3$ is less than $\rho_i$ = 915 kgm$^{-3}$, used [from Kurtz et al, 2012] to derive SIT from OIB and the ice density 925kg/m$^3$, used from [Kwok and Cunningham, 2008] to derive SIT from ICESat, or the $\rho_i$ = 900 kgm$^{-3}$, used for SIT retrieval from ERS and ENVISAT RA [Connor, et al, 2009], which may lead to underestimation of SIT in comparison with SIT derived from OIB, ICESat, or ERS1, 2 and ENVISAT, if a (point) fixed ice density is used for the same conditions (freeboard, snow depth, ice type and temperature).

Surface [Wadhams et al, 1992] and theoretical studies [Kovacs, 1996] confirm dependence of $\rho_i$ on ice type and $h_{fi}$. Surface [Ackley, et al, 1976], satellite [Connor et al, 2009] observations and sensitivity study (in the next sections) confirm that use of not correct, fixed ice densities to convert F to SIT (using Equation 5) may introduce significant error in dependence on ice type and freeboard.

Ackley et al, (1976), carried out a point-by-point isostatic analysis of MY floes, which have been profiled by drilling and tested few algorithms for freeboard-to-draft conversion: i) a simple point isostatic model, using an estimated mean ice density and ii) a variable, free-board dependent (FD) density model based on regression dependence of ice density ($\rho_i$) on "effective" ice free-board $h_{fie}$ (m). Based on surface observations (drilling) of $h_{fi}$ and d, Ackley et al, (1976), estimated a linear relationship between $\rho_i$ and "effective" ice free-board $h_{fie}$ (m):

$$\rho_{iMYAK} = -a_1 h_{fie} + b \qquad (12)$$

where $h_{fie}$ depends on sea ice freeboard $h_{fi}$, snow depth $h_s$, density $\rho_s$ and the mean MY ice density, $\rho_{iMYmean}$:

$$h_{fieMY} = h_{fi} + (h_s \rho_s / \rho_{iMYmean}) \qquad (13)$$

where $a_1=194$, $b=948$ are estimated for $\rho_w=1020 kg/m^3$ [Wadhams, et al 1992, Ackley et al, 1976].

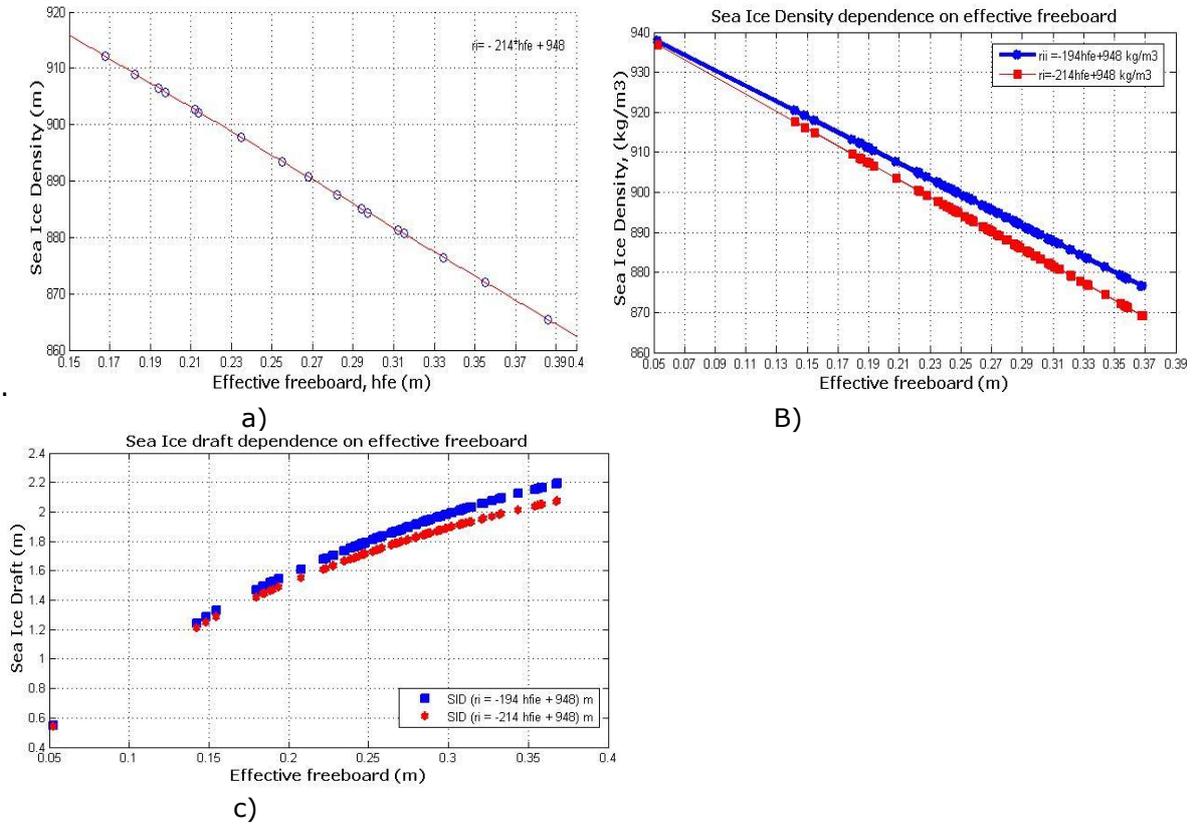

Figure 2 Freeboard dependent ice densities and SID.

By comparison with in-situ (drilling) observations, Ackley et al, 1976 concluded that the FD ice density model provides more accurate results for F to SID conversion compared with use of point (fixed) sea ice density. Assuming hydrostatic equilibrium and applying sensitivity analyses, the coefficients (a,b) in Equation 12 have been updated (a=214, b=948 ) to satisfy FD ice density over MYI :

$$\rho_{iMYFD} = -a\, h_{fie} + b \qquad (14)$$

with coefficients a=214, b=948, $h_{fie}$, calculated by Equation 13 (in the range 0.18m, 0.37m) for wide range of F (0.177m to 0.367m), $\rho_w$ =1024 kg/m$^3$, $\rho_{iMYmean}$ =882kg/m$^3$ , $\rho_s$ ( from 260 kg/m$^3$ to 360kg/m$^3$), $h_s$ (from 0.01m to 0.38m), corresponding to snow depth and density range observed in the Arctic and confirmed by WC for all winter months and different locations. The FD ice density (Figure 2/a), satisfying Equation 14, has been validated and compared with $\rho_{iMYAK}$ (Figure 2/b) for wide range of $h_{fi}$ corresponding to MYI. The small over estimation of $\rho_{iMYAK}$ compared with $\rho_{iMYFD}$, is due to different water density, used from Ackley, et al, (1976), which impacts the slope of the retrieved $\rho_{iMYAK}$ and leads to overestimation of SID (Figure 2/c). The FD ice density algorithms $\rho_{iMYFD}$ (Equation 14, for $h_{fie}$ in the range 0.18m/0.37m) has been extended over FYI and MYI with $h_{fie}$ >0.37m and has been validated using Kovacs (1996) observations.

Two algorithms have been applied and validated for $\rho_{iFY}$ over FYI (or effective freeboard <0.18m): i) sea ice density over FYI (for $h_{fie}$<0.18m) was set to the mean FYI ice density $\rho_{iFYC}$ = 910kg/m$^3$, according to Kovacs, 1996; ii) the FYI FD ice density, $\rho_{iFYFD}$, has been retrieved by the following regression:

$$\rho_{iFYFD} = a\, h_{fieFY} + b \qquad (15)$$

Where
$$h_{fieFY} = h_{fi} + (h_s\rho_s/\rho_{iFYmean}) \tag{16}$$

with a=-95.05, b=930.4 and $\rho_{iFYmean}$ =910kg/m$^3$, providing the best fit to $\rho_{iFYKO}$, observed by Kovacs over FYI. The MYI ice density, $\rho_{iMYE}$, has been extended (Equation 14) for $h_{fe}$>0.37m with a=36.54, b=903.7, RMS=0.518, satisfying the observed $\rho_{iMYKO}$ and SIT by Kovacs, 1996 for MYI. The accuracy of the FD ice density depends mostly on the accuracy of the $h_{fi}$, and is summarised in Table 2 for 0.097m< $h_{fie}$ < 0.53m.

Table 2. Impact of sea ice freeboard uncertainty $\sigma_{hi} = \pm0.03$m ($h_{fi} \pm0.03$m) on accuracy of the retrieved FD ice density, $\rho_{iFD}(h_{fi})$, for 0.0m< $h_{fie}$ < 1.2m

| Sea ice freeboard ($h_{fi} \pm0.03$m) | FD Ice density uncertainty, $\rho_{iFD}\pm\sigma_{\rho iFD}$ (kg/m$^3$) |
|---|---|
| 0.00 m< $h_{fi}$ < 0.18m  FYI | $\sigma_{\rho iFD} = \pm2.85$ kg/m$^3$ |
| 0.18 m< $h_{fi}$           MYI | $\sigma_{\rho iFD} = \pm3.45$ kg/m$^3$ |

The uncertainty of the freeboard depended ice density depends on uncertainties of the sea ice freeboard (e.g. $h_{fi} \pm0.03$m will introduce sea ice density uncertainty, $\sigma_{\rho iMYFD} = \pm3.45$ kg/m$^3$ for MYI, or $\sigma_{\rho iFYFD} = \pm2.85$ kg/m$^3\sigma$ for FYI), which will decrease if the accuracy of the sea ice freeboard increase. The retrieved regression coefficients (a,b) (for Equations 14/15) satisfy [Kovacs, 1996, Ackley, 1876] observations of $\rho_i$, equation for hydrostatic equilibrium (for $h_{fei}$ from 0.001m to over 1.2m), covering wide range of $h_{fi}$, $h_s$, $\rho_s$ and are validated by comparison of derived SID(RA, A(FD)) from RA with SID(ULS). The total ice density, $\rho_i$, considering the contribution of the ice densities from FYI and MYI, is calculated as a function of the ice densities of FYI and MYI, considering their fractions:

$$\rho_{iFD}=fr(FY)\rho_{iFY} +(1-fr(FY))\rho_{iMYFD} \quad (kg/m^3) \tag{17}$$

where the fraction fr(FY) of the FYI cover from the RA averaged area is taken from OSI –SAF, $\rho_{iFY}$ =910 kg/m$^3$ (A3,FD), or $\rho_{iFYFD}$ is calculated by Equation 15-16 (A5,FD2), $\rho_{iMYFD}$ is calculated by Equation 14 with a=214, b=948, for 0.18m< $h_{fie}$ >0.37m and a=36.54, b=903.7 for $h_{fe}$>0.37m. The freeboard, snow depth and ice type dependent sea ice density (calculated by Equations 17) is inserted in Equation 5 for conversion of RA freeboard to SID. The SID and regularisation constants (a, b), determining weighted contribution of ice freeboard and snow depth in the derived FD ice density over FYI and MYI have been validated with independent SID(ULS) observations in different locations within 12 years. The high accuracy of the freeboard depended algorithm is confirmed with minimum RMS and bias (up to $\varepsilon$=0.001m) of the retrieved SID from RA, using FD algorithm for ice density, when compared with collocated SID (ULS).

## 2.2. Sensitivity of retrieved SIT on input parameter variability

The purpose of sensitivity analyses is to answer the following questions: i) what are the uncertainties in the retrieved SIT, due to uncertainties of the ice density for MYI and FYI and different snow depths, keeping the same freeboard (Figure 3)?; ii) what will be the error in calculated SIT if we use constant ice density, or wrong ice density?; iii) what will be the error, introduced in SIT for different freeboards (Figure 3/b) due to uncertainties of the input sea ice density and freeboard for typical snow depth (30cm)?; iv) what will be the error in the retrieved SIT if snow depth is underestimated, or we consider 0.5$h_s$(WC) over FYI; v) what is the impact of snow and water density on retrieved SIT? Vi) what is the impact of freeboard depended ice density on accuracy of the retrieved SIT?
The snow density from WC and constant water density $\rho_i$=1025kg/m$^3$ are used for sensitivity analyses (Figure 3). Wider range of sea ice density (720- 950kg/m$^3$) is applied because this is the recommended ice density from different authors and allows to investigate the ice density impact in any smaller range (e.g. 860-940kg/m$^3$). The sensitivity of SIT (calculated by Equation 5, assuming hydrostatic equilibrium) to $\rho_i$, $h_s$ and $h_{fi}$ is shown on Figure 3.
The impact of uncertainties of fixed sea ice density is illustrated with Figure 3/a,b. Applying equation for hydrostatic equilibrium, using fixed ice density, $916.7\pm35.7 kg/m^3$ (used in Algorithm A2), and snow depth 0.05m over FYI will produce SIT=1.096m with uncertainties $\pm0.2717m$ for $h_{fi}=0.1m$ and the SIT uncertainty will increase, when snow depth, density and ice freeboard increase (Figure 3/a).

The impact of a wrong, constant ice density on accuracy of the retrieved SIT is shown on Figure 3/a. If the area with MYI is misclassified as a FYI, use of a constant (wrong) FY ice density ($916.7 kg/m^3$) instead of MYI density ($882 kg/m^3$) will lead to SIT=$4.4086\pm1.093m$ (for $h_{fi}=0.45m$) instead of SIT=$3.3388\pm0.4626m$, overestimating SIT with $1.0698\pm0.63m$ for the same snow depth ($h_s=0.3m$, Figure 3/b). The retrieved SIT by Equation 5 for sea ice freeboard 0.27m and ice density $882 kg/m^3$ will be underestimated in the range 50cm - 80cm, for snow depth in the range 0-40cm compared with SIT calculated by A1 algorithm using ice density $900 kg/m^3$ and the residual will depend on ice type and snow depth (Figure 3/a). Ice density is the most important variable, impacting the accuracy of the retrieved SIT from RA and use of not accurate ice density may lead to up to 1.6m bias in the retrieved SIT for $h_{fi}=0.27m$ in absence of snow depth. To avoid wrong estimate of SIT due to use of inaccurate fixed ice density, not accounting for ice type, $h_s$ and $h_{fi}$, a freeboard and ice type depended ice density, $\rho_{iFD}$, (Equations, 15-17, Figure 2) is necessary to insert in Equation 5, allowing to calculate correct SIT in dependence on sea ice type and corresponding ice density.

The impact of the uncertainty of the freeboard on accuracy of the retrieved SIT is illustrated with Figure 3/b when fixed ice density is used. One can see that sea ice freeboard uncertainties ($0.2\pm0.03m$) will lead to uncertainties in the SIT retrieved from RA (in the range $2.277\pm0.238m$ when fixed $\rho_i=900 kg/m^3$, $\rho_w=1030 kg/m^3$, $\rho_s=300 kg/m^3$ and $h_s=0.3m$ are used) and the uncertainties of SIT will increase when the freeboard increase if not accurate ice density is used. The freeboard uncertainty impact on retrieved SIT from RA will decrease when FD ice density is applied (Figure 3/b), providing lower ice density for MYI (over high freeboard) and higher values for ice density over FYI (with small freeboard). Thus, the freeboard depended ice density minimise the impact of the uncertainties in the sea ice freeboard and snow depth.

The impact of the assumption for half snow depth over FYI is illustrated with Figure 3/a, Figure 4 and Table 3. SIT will be underestimated with 0.46m to 0.53m for $h_{fi}$ from 0.1m to 0.2m and the underestimation of SIT will be higher ($\varepsilon=-0.5296m$ for $\rho_i=916.7 kg/m^3$, A2, CryoSat2) compared with A1 ($\varepsilon=-0.462m$ for $\rho_i=900 kg/m^3$, A1) if a wrong assumption for half snow depth over FYI is applied (Table 3). One can see from Figure 4 that presence of 40cm snow depth, not considered in the calculations of SIT, may lead to underestimation of SIT between 0.41m to 1.6m in dependence of the ice density, used to calculate SIT by Equation 5, which confirms the higher impact of ice density on retrieved SIT than the snow depth. The impact of snow depth on retrieved SIT from RA increases when the snow depth increase (Figure 3/a, Table 5) and depends on ice density. In absence of snow ($h_s=0m$) the uncertainty due to use of not accurate fixed ice density is up to 0.7784m and it is increased only with 0.088m, due to 10cm impact of snow depth, which confirms that the ice density is the most important variable, which impacts the accuracy of the retrieved SIT from RA. Only a freeboard depended ice density, considering the ice type and snow depth along the RA track will minimise the impact of uncertainties in the retrieved SIT from RA by inserting $\rho_i$, in Equation 5, calculated as a function of effective freeboard (Figure 2).

Table 3 Impact of the assumption for half snow depth over FYI, sea ice freeboard and ice density for ($\rho_i=900 kg/m^3$ and $\rho_i=882 kg/m^3$), $\rho_s=300 kg/m^3$, $\rho_w=1300 kg/m^3$

| $\rho_i = 900 kg/m^3$ | | $\rho_i = 900 kg/m^3$ | | $\rho_i = 916.7 kg/m^3$ | | $\rho_i = 916.7 kg/m^3$ | |
|---|---|---|---|---|---|---|---|
| $h_{fi}=0.1m$ | | $h_{fi}=0.2m$ | | $h_{fi}=0.1m$ | | $h_{fi}=0.2m$ | |
| $h_s=0.4m$ | $h_s=0.2m$ | $h_s=0.4m$ | $h_s=0.2m$ | $h_s=0.4m$ | $h_s=0.2m$ | $h_s=0.4m$ | $h_s=0.2m$ |
| $h_i(m)=1.7154$ | 1.2538 | 2.508 | 2.046 | 1.9682 | 1.4387 | 2.8773 | 2.3477 |
| $h_i(h_s=0.2)- h_i(h_s=0.4)=$ $\varepsilon=-0.4612(m)$ | | $\varepsilon=-0.462(m)$ | | $\varepsilon=-0.529(m)$ | | $\varepsilon=-0.5296(m)$ | |

Table 4. Impact of snow density and $\rho_w$ on SIT, using the Equation of hydrostatic equilibrium and $\rho_i=900 kg/m^3$

| $h_s=0.3m$, $h_{fi}=0.27m$, $\rho_w=1030 kg/m^3$ | | $h_s=0.3m$, $h_{fi}=0.27m$, $\rho_w=1024 kg/m^3$ | |
|---|---|---|---|
| $\rho_s$ kg/m$^3$ | $h_i$ (m) | $\rho_s$ kg/m$^3$ | $h_i$ (m) |

| 260 | 2.739 | 260 | 2.869 |
| 300 | 2.83 | 300 | 2.956 |
| 360 | 2.97 | 360 | 3.1 |

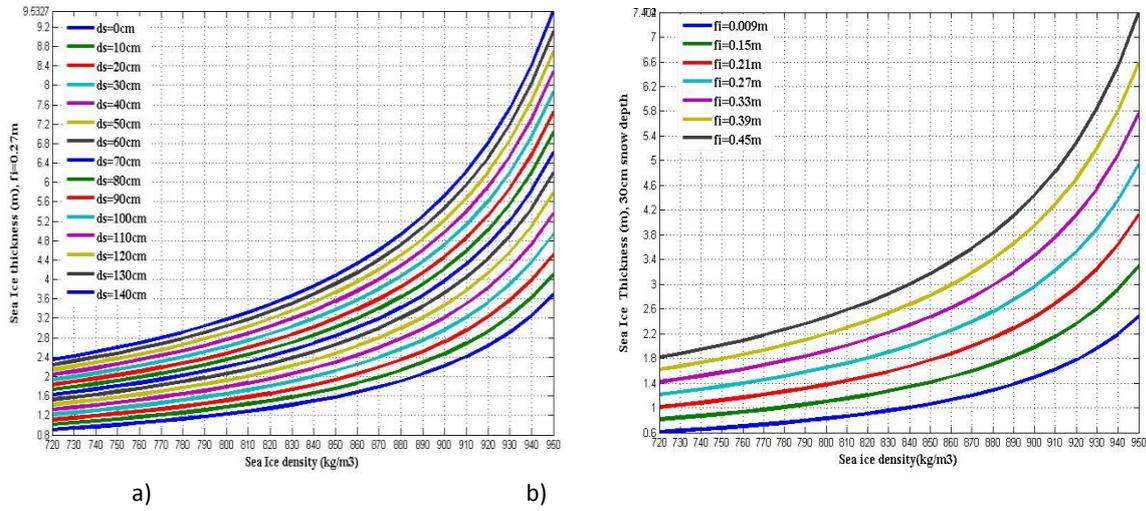

Figure 3. a) Sensitivity of SIT to sea ice density and snow depth; b) sensitivity of the retrieved SIT on ice density for snow depth 30cm and free-board change from 0.09m to 0.45m.

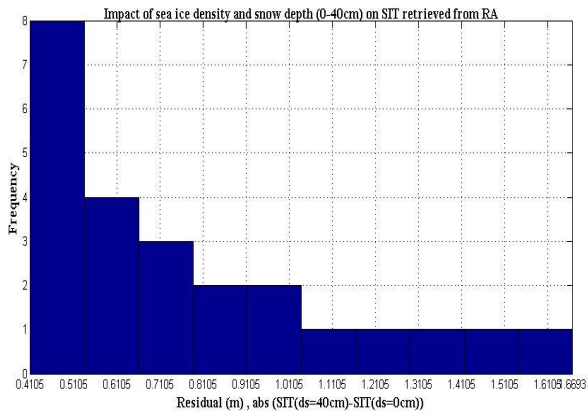

Figure 4. Impact of 40 cm decrease in snow depth on accuracy of SIT if a fixed ice density in the range 720 to 950kg/m$^3$ is used to calculate SIT.

Table 5. Standard deviation of SIT for snow depth from 0 to 140cm and ice density from 720 to 950kg/m$^3$.

| $h_s$(cm) | 0 | 10 | 20 | 30 | 40 | 50 | 60 | 70 |
|---|---|---|---|---|---|---|---|---|
| Std (m) | 0.7784 | 0.8664 | 0.9545 | 1.0425 | 1.1305 | 1.2186 | 1.3066 | 1.3946 |
| Var (m) | 0.6059 | 0.7507 | 0.9110 | 1.0868 | 1.2781 | 1.4849 | 1.7072 | 1.9450 |
| $h_s$(cm) | 80 | 90 | 100 | 110 | 120 | 130 | 140 | Mean |
| Std (m) | 1.4827 | 1.5707 | 1.6587 | 1.7468 | 1.8348 | 1.9228 | 2.0109 | |
| Var (m) | 2.1983 | 2.4671 | 2.7514 | 3.0512 | 3.3665 | 3.6973 | 4.0436 | 2.0897 |

Smaller FD ice densities (calculated by Equation 15-17), corresponding to MYI with high $h_s$, $h_{fi}$ will lead to decreased uncertainty due to snow depth and sea ice freeboard (Figure 3/a, b), which is not the case when a fixed ice density is used.

Snow and water density also impacts the accuracy of the retrieved SIT (Table 4). SIT will be underestimated with 0.15m (from 3.12m to 2.97m) applying A1 ($\rho_i$ = 900 kg/m$^3$, $\rho_w$=1030kg/m$^3$, $h_{fi}$=0.3m, $h_s$=0.3m) if we use $\rho_s$ =260kg/m$^3$ (Used from OIB2009) instead of $\rho_s$ =320kg/m$^3$ (Used from OIB2010). The bias ($\varepsilon$=$h_i$ ($\rho_s$=360kg/m$^3$)-$h_i$($\rho_s$=260kg/m$^3$)) will increase to 0.24m (for $\rho_i$ =900 kg/m$^3$, $\rho_w$=1030kg/m$^3$, $h_{fi}$=0.27m, $h_s$=0.3m) if we use $\rho_s$=360kg/m$^3$ instead of $\rho_s$ =260kg/m$^3$ (Table 4). Use of higher water density ($\rho_w$ =1030kg/m$^3$, A1, A2 (CryoSat)), instead of $\rho_w$ =1024kg/m$^3$ will lead to underestimation of SIT with about 0.13m for the same $\rho_s$, $\rho_i$, $h_s$, $h_i$ (Table 4). The sensitivity analysis confirms:

i) The sea ice density has the greatest impact on uncertainties of the retrieved SIT from RA, leading to $\sigma(\rho_i)$=0.7784m, in absence of snow depth if not accurate fixed ice density is used, which is about 10 times higher than the uncertainty introduced from 10cm uncertainty in snow depth;

ii) Use of a wrong fixed ice density could introduce up to 1.6m bias in the retrieved SIT, when the equation for hydrostatic equilibrium is applied to calculate ;

iii) Uncertainty $\sigma$= $\pm$0.03m in the sea ice freeboard (for $h_{fi}$ =0.2m $\pm$0.03m) will lead to uncertainties in the SIT retrieved from RA ($\sigma(h_i)$=0.238m, $h_i$=2.277$\pm$0.238m, when fixed $\rho_i$ =900kg/m$^3$, $\rho_w$ =1030kg/m$^3$, $\rho_s$ =300kg/m$^3$ and $h_s$ =0.3m are used) and the freeboard impact on retrieved SIT from RA will decrease about 6 times ($\sigma(h_i)$=0.043m for $h_i$ =0.1m, $\rho_i$ =900kg/m$^3$, $\rho_w$ =1024kg/m$^3$, $\rho_s$ =300kg/m$^3$ and $h_s$ =0.3m) when FD ice density is applied ;

iv) The assumption of the half snow depth over FYI will always lead to underestimation of the $h_i$ if the equation for hydrostatic equilibrium is applied and the bias will be higher for A2 (CryoSat2, using fixed ice density, $\rho_{iFY}$ =916.7kg/m$^3$) than for ice density, $\rho_i$ =900kg/m$^3$.

v) Use of a wrong snow density ($\rho_s$ =260kg/m$^3$ instead of $\rho_s$ =360kg/m$^3$ ) may introduce higher uncertainties ($\sigma(\rho_s)$=0.24m) in the retrieved SIT from RA than the sea ice freeboard ($\sigma(h_{fi})$=0.238m). Use of $\rho_w$ =1030kg/m$^3$ (Laxon et al 2012) instead of $\rho_w$ =1024kg/m$^3$ (used from ICESat ) will lead to 0.1-0.13m underestimation of SIT when the hydrostatic equation is applied.

vi) The uncertainty due to ice density impact will decrease from $\sigma(\rho_i)$=0.7784m to 0.0239m in absence of snow, which is 32.43 times improvement in the accuracy of the retrieved SIT from RA if a FD algorithm is applied.

## 2.3. Algorithms for SIT retrieval from RA data

Two main groups of algorithms for conversion of RA measured freeboard to SIT have been tested: i) Algorithms based on hydrostatic equilibrium, using different input variables of ice density and snow depth; ii) Empirical regression equations. All algorithms used until now have been validated and 3 new algorithms have been included to test 2 hypothesis: i) impact of freeboard dependent ice density of MYI and FYI on accuracy of the retrieved SID from RA; ii) impact of assumption for half WC snow depth over FYI when fixed ice density (900kg/m$^3$ ) is applied.

**Algorithms based on Hydrostatic equilibrium**

Two types of algorithm for free-board to SIT conversion have been selected for validation: i) using fixed ice densities; ii) freeboard dependent (FD) algorithm.

**Algorithm 1 (A1) - fixed ice density and snow depth from Warren climatology**

The freeboard, derived from RA, is converted to SIT, using hydrostatic equilibrium (Equation 5) with ice density ($\rho_i$ = 900kg/m$^3$), $\rho_w$=1030 kg/m$^3$, snow depth and density, retrieved by WC. SIT, SID and F from RA are collocated with ULS data for algorithm comparison and selection.

**Algorithm 2 (A2) – fixed ice densities for FYI and MYI ice and snow depth from Warren climatology over MYI, half WC snow depth over FYI, (CryoSat-2)**

The freeboard, derived from RA, is converted to SIT, using hydrostatic equilibrium (Equation 5), where ice density is calculated by:

$$\rho_i = fr(FY)\ 916.7 + (1-fr(FY))882 \quad (kg/m^3) \tag{18}$$

with fixed ice densities: $\rho_{ify}$ = 916.7 kgm$^{-3}$ for FY ice and $\rho_{iMY}$ =882 kgm$^{-3}$ for MYI, $\rho_w$=1030 kg/m$^3$ and snow depth is calculated by:

$$h_s = 0.5fr(FY)*h_s\ (WC)+(1-fr(FY))h_s\ (WC) \tag{19}$$

where $h_s(WC)$ is the snow depth for MYI from WC and half of the snow depth from WC ($h_{sfy}=0.5h_s\ (WC)$) is used for FYI. The fraction, $fr(FY)$, of FYI from the RA averaged area is available from OSI –SAF from 2005.
Assumption of $h_{sfy}=0.5h_s\ (WC)$ is based only on limited high resolution observations of snow depth, using OIB radar for April, 2009 [Farrel et al, 2012]. The equation for hydrostatic equilibrium and sensitivity analyses show that decrease 2 times of snow depth will lead to essential underestimation of SIT and SID derived from RA. If information for FYI ($fr(FY)=0$) is not available or over MYI the SIT and SID are calculated from RA for fixed $\rho_i=882$ kgm$^{-3}$, which is lower than the ice densities, used for A1 ($\rho_i=900$ kgm$^{-3}$), from ATM/OIB ($\rho_i=915$ kgm$^{-3}$) and from ICESat ($\rho_i=925$ kgm$^{-3}$) and any comparison may lead to misleading results.

**Algorithm 3 (MFD). MY Freeboard dependent Algorithm (FD)**

SIT is calculated as a function of freeboard, derived from RA, using hydrostatic equilibrium (Equation 5), where the freeboard dependent ice density over MYI is estimated by Equation 14 ($\rho_{iMY}$ = -a$h_{fie}$ + b, with a=214, b=948, $\rho_{iMYmean}$= 882kg/m$^3$ and the total ice density is calculated by Equation 17, where $\rho_{iFY}$=910kg/m$^3$ for FYI, snow depth and density are calculated from WC and water density is 1024kg/m$^3$.

**Algorithm 4 (A4). Fixed ice densities and snow depth over FYI half of the $h_s$ from Warren climatology.**

The assumption [Laxon et al, 2012] of half snow depth over FYI, based on few tracks of OIB in 2009, has been tested applying sensitivity analyses and Algorithm A4, where SIT is calculated from RA, using hydrostatic equilibrium (Equation 5) with fixed ice densities: $\rho_i$ = 900 kgm$^{-3}$, $\rho_w$ =1030 kg/m$^3$ as (in A1 Algorithm), snow density is from WC and snow depth is 0.5$h_s$(WC) over FYI. The sensitivity analyses and validation with independent ULS data confirm that SIT and SID, will be always underestimated if we assume half snow depth over FYI.

**Algorithm 5 (FD). Freeboard dependent Algorithm (FD2)**

SIT is calculated as a function of freeboard, derived from RA, using hydrostatic equilibrium (Equation 5), with inserted a freeboard dependent ice density over MYI, estimated by Equation 14-17 ( $\rho_{iFY}$ = -a $h_{fie}$ + b), where:

a=214, b=948, for 0.18m <$h_{fie}$ <0.37m, $\rho_{iMYmean}$= 882kg/m$^3$ , over MYI
a=36.54, b=903.7, for $h_{fie}$ >0.37m, $\rho_{iMYmean}$= 882kg/m$^3$ , over MYI
a=95, b=930.4 , for $h_{fe}$<0.18m, $\rho_{iFYmean}$=910kg/m$^3$ , over FYI

and the total ice density is calculated by Equation 17, where, snow depth and density are calculated from WC and water density is 1024kg/m$^3$.

**Empirical regression algorithms (A3, W) to retrieve SIT from RA**

**Algorithm 6 (A5) – empirical relationship between freeboard and SIT by Alexandrov et al. 2010**

Applying regression analyses, $h_i$ is approximated as a function of $h_{fi}$ for FYI and MYI [Alexandrov et all, 2010]. The regression equations are retrieved for sea ice density (MYI, $\rho_i$ =882$\pm$23kg/m$^3$, FYI, 916.7+35.7 kg/m$^3$), water density, $\rho_w$ =1025+0.5 kg/m$^3$ and snow density $\rho_s$ =324 kg/m$^3$$\pm$50 kg/m$^3$ (FYI), $\rho_s$ =320$\pm$20 kg/m$^3$ (MYI):

$h_i = 9.46\ h_{fi} + 0.15$ for FYI (20)

$h_i = 6.24\ h_{fi} + 1.07$ for MYI (21)

The equations have been retrieved for input variables for $h_s$, $\rho_s$, $\rho_i$, $\rho_w$ from Sever expedition, which restricts the application of Equations 20, 21.

**Algorithm 7 (W) – empirical relationship between freeboard and SIT by Wadhams 2000**

Based on airborne laser altimeter and submarine sonar data Wadhams has found an empirical relation between the freeboard ($h_{fi}$) and ice draft of thick MY ice north of Greenland:

$h_i = 9.04\ h_{fi}$ (22)

Considering that the SIT measured from laser altimetry depends on snow depth and density (if present) Equation 22 is snow depth dependent also and depends on local snow depth and density, which makes this algorithm with restricted local application.

## 2.4. Uncertainty analysis of retrieved SIT

Only the uncertainties, associated with conversion of freeboard to thickness (using Equation 5 for hydrostatic equilibrium) will be analysed. Assuming that the input variables in Equation 5 are uncorrelated, the uncertainty ($\sigma_{hi}$) of the retrieved thickness, hi, from the freeboard, measured from RA, will depend on propagated uncertainties of the input variables [Kwok, 2010]:

$\sigma^2_{hi} = \sigma^2_{hfi}(dh_i/(dh_{fi}))^2 + \sigma^2_{hs}(dh_i/(dh_s))^2 + \sigma^2_{\rho s}(dh_i/(d\rho_s))^2 + \sigma^2_{\rho i}(dh_i/(\rho_i))^2 + \sigma^2_{\rho w}(dh_i/(\rho_w))^2$ (23)

where $\sigma_{hi}$, $\sigma_{hfi}$ and $\sigma_{hs}$ are the uncertainties of the SIT, F and $h_s$, $\sigma_{\rho s}$, $\sigma_{\rho w}$, $\sigma_{\rho i}$ are the uncertainties in the snow, water and ice densities. The mean values and uncertainties of all 5 algorithms (A1-A5), (based on Equation 5) are listed in Table 6.

Table 6 Uncertainties of the main variables, contributing to SIT retrieval using the Equation for hydrostatic equilibrium, applying different algorithms

| Alg. | A1 | A1 | A2/FY | A2/FY | A2/MY | A2/MY | A3/FD | A3/FD | A4/FY | A4/FY | A5/FY | A5/FY | A5/MY | A5/MY |
|---|---|---|---|---|---|---|---|---|---|---|---|---|---|---|
| Variable | Mean | σ | mean | σ | mean | σ | Mean | σ | mean | σ | mean | σ | mean | σ |
| $h_{fi}$ (m) | 0.3 | ±0.03 | 0.3 | ±0.03 | 0.3 | ±0.03 | 0.3 | ±0.03 | 0.3 | ±0.03 | 0.3 | ±0.03 | 0.3 | ±0.03 |
| $h_s$(cm) | 29.1 | ±0.075 | 14.5 | ±0.075 | 29.1 | ±0.075 | 29.1 | ±0.075 | 14.5 | ±0.075 | 29 | ±0.075 | 29 | ±0.075 |
| $\rho_i$ kg/$^3$ | 900 | ±50 | 916.7 | ±35.7 | 882 | ±23 | 900 | ±3.45 | 900 | ±50 | 915 | ±2.8 | 900 | ±3.45 |
| $\rho_s$ kg/$^3$ | 295 | ±4.4 | 295 | ±4.4 | 295 | ±4.4 | 295 | ±4.4 | 295 | ±4.4 | 295 | ±4.4 | 295 | ±4.4 |
| $\rho_w$ km$^{3*}$ | 1030 | ±6 | 1030 | ±6 | 1030 | ±6 | 1024 | ±0.2 | 1030 | ±6 | 1024 | ±0.2 | 1024 | ±0.2 |

The mean freeboard ($h_{fi}$ =0.3m), from RA in the Arctic with uncertainty $\sigma_{hfi}$ =±0.03m [Connor et al, 2009] and the mean snow density, $\rho_s$=295 kg/m$^3$ with $\sigma_{\rho s}$= ±4.4kg/m$^3$ [Radionov et al, 1997, Warren,1999] have been applied for all algorithms. The mean snow depth 29.13cm with uncertainty $\sigma_{hs}$= ±0.075cm from WC [Warren, et al, 1999] in winter months have been applied for all algorithms, excluding A2(FYI) and A4(FYI), assuming half snow depth over FYI, where mean snow depth of 14.5cm is applied. Ice density for FYI $\rho_i$=916.7kgm$^{-3}$ with uncertainty ±35.7 kgm$^{-3}$ and MYI density $\rho_i$ = 882kg/m$^3$ with uncertainty ±23kg/m$^3$[Alexandrov, et al, 2010] are used in A2 algorithm. Water density depends on water temperature and salinity. Water density $\rho_\omega$ =1030kg/m$^3$ (biased with 6 kg/m$^3$) than the measured one in Beaufort Sea (1024kg/m$^3$) have been used for Algorithms A1, A2 and A4 and the water density $\rho_\omega$=1024kg/m$^3$, have been used for FD algorithms (A3, A5). The accuracy of $\rho_i$=900kg/m$^3$ for A1 and A4 is not known and 900±50kg/m$^3$ (in the range 850-950 kg/m$^3$) is assumed. Considering the strong dependence of sea ice density on ice type, temperature and freeboard, Algorithms A1 and A4 are with largest uncertainties due to sea ice density impact and the A3 and A5 (FD) are with smallest uncertainty $\sigma_{\rho iMYFD}= \pm3.45 kg/m^3$ $\sigma_{\rho iFYFD}= \pm2.8 kg/m^3$ $kg/m^3$ due to $\pm0.03m$ variations in $h_{fi}$. Because the uncertainties of the input variables in Equation 5 and Table 6 are in different units, the percentage contribution of uncertainty ($\sigma\% =\sigma*100/mean$) of input variables for all 5 Algorithms and the total uncertainties are given in Table 7.

Table 7. Percentage contribution of uncertainties of input variables and the total uncertainties ($\sigma\%$) of the calculated SIT, using the Equation for hydrostatic equilibrium

| Alg. | A1 | A2/FY | A2/MY | A3(FD) | A4/FY | A5/FY | A5/MY(FD2) |
|---|---|---|---|---|---|---|---|
| Variable | $\pm \sigma\%$ | | | | | | |
| $h_{fi\,i}$ | $\pm 10\%$ | | | | | | |
| $h_{s\,i}$ | $\pm 0.26\%$ | $\pm 0.52\%$ | $\pm 0.26\%$ | $\pm 0.26\%$ | $\pm 0.52\%$ | $\pm 0.26\%$ | |
| $\rho_i$ | $\pm 5.6\%$ | $\pm 3.89\%$ | $\pm 2.6\%$ | $\pm 0.38\%$ | $\pm 5.6\%$ | $\pm 0.3\%$ (FYI); $\pm 0.38\%$ MYI | |
| $\rho_w$ | $\pm 0.58\%$ | $\pm 0.58\%$ | $\pm 0.58\%$ | $\pm 0.0195\%$ | $\pm 0.58\%$ | $\pm 0.0195\%$ | |
| $\rho_s$ | $\pm 1.49\%$ | | | | | | |
| $\sigma\%$ total | 17.93% | 16.48% | 13.93% | 12.1% | 18.2% | 12% | 12.1 |

One can see that the smallest uncertainties in the retrieved SIT are observed for A(5,FD2), (only 2% not considering the impact of the freeboard uncertainty). The impact of snow depth on uncertainties is negligible (less than 0.5% for all algorithms, except for A2 and A4 over FYI, where $\sigma_{hs}=0.52\%$ when half snow depth over FYI is assumed). Sea ice freeboard accuracy and ice density are the most important factors contributing to the uncertainty. The uncertainty 5.6%, calculated for fixed ice density 900kg/m$^3$, assuming uncertainty $\sigma_{\rho t}=\pm50kg/m^3$ is the smallest sea ice uncertainty for fixed ice density 900kg/m$^3$. Considering [Alexandrov et al, 2010 and Tinco and Frederico, 1996] the ice density range could be wider (720-950kg/m$^3$), introducing sea ice thickness uncertainty ($\sigma_{hi}=0.78m$, for $h_s=0$, Table 5) in the retrieved SIT from RA in dependence of ice type and freeboard and absence of snow. The uncertainty of sea ice density is reduced to 0.3% when FD ice density is applied for freeboard accuracy 10%, which confirms the advantage of the FD ice density.

## 3. Algorithm comparison

The accuracy of the retrieved SID from RA, applying different algorithms, has been also estimated by comparison of SID, derived from RA, with SID retrieved from collocated ULS data with known accuracy, applying different algorithms (A1-A7) and statistical analyses.
The equations and Algorithms applied to calculate SID(RA) are shown on Figure 5.
Seven algorithms have been compared, using collocated ULS data from moored upward looking sonar (ULS) and ULS on submarine (averaged over 50km transect along the submarine track): i) Fixed ice density and snow depth and density from WC, assuming hydrostatic equilibrium [Connor et al, 2009]; ii) CryoSat2 algorithm based on hydrostatic equilibrium equation, using fixed values for ice density for FYI and MYI and half of the snow depth from WC over FYI [Laxon, 2012]; iii) Freeboard depended (FD) algorithm for MY sea ice density, where the snow depth and density are from WC, assuming hydrostatic equilibrium; iv) Fixed ice density, half snow depth from WC over FYI and snow density from WC; v) Extended Freeboard depended (FD2) algorithm for FYI and MYI, assuming hydrostatic equilibrium; vi) empirical algorithm [Alexandrov et al, 2011]; vii) empirical algorithm [Wadhams (1992, 2000)].
An algorithm for conversion of F to SIT is selected based on a Statistic, calculated for SID, derived from RA and ULS, collocated data in different locations of the Arctic in the period 1996 to 2008. The following statistical variables are calculated for SID, derived from RA and ULS: minimum (min), maximum (max), mean (m), median, mode, mean residual ($\epsilon$=SID(RA, Ai)-SID(ULS)), correlation coefficient (r), best fit coefficients and RMS error, applying graphical and histogram analyses. Error bar of the mean SID, calculated for ULS and RA, using algorithms A1-A7, with corresponding deviations ($2\sigma$) is also plotted.

The impact of the uncertainties, contributing to SID retrieval is minimized by the mean residual, $\varepsilon$, which makes minimum $\varepsilon$ one of the criteria for algorithms comparison. Algorithm with minimum $\varepsilon$ and min RMSE for all collocated RA and ULS data (from 1996 to 2008) is selected. Only collocated independent observations of SID from ULS (from NSIDC and BGEP Experiment) with known accuracy have been selected for algorithm comparison. ULS data with high bias, low correlation, or not complete sections of 50km cover have been excluded from validation because the assessment of algorithm accuracy relay on accuracy of ULS data used for comparison. SID(ULS, 1994) have not been used for algorithm validation due to very low correlation (0.237) of SID(ULS) and SID(RA). The SIT, calculated from OIB observations in 2009 and 2010 [Kurtz et al 2012] and snow depth from OIB/radar [Farell et al, 2012] have been also excluded on this study from the data sets used for Algorithm selection due to use of different input variables to retrieve SIT from OIB and SIT from RA with unknown uncertainties[Kurtz et al, 2012, Cavalieri et al, 2012].

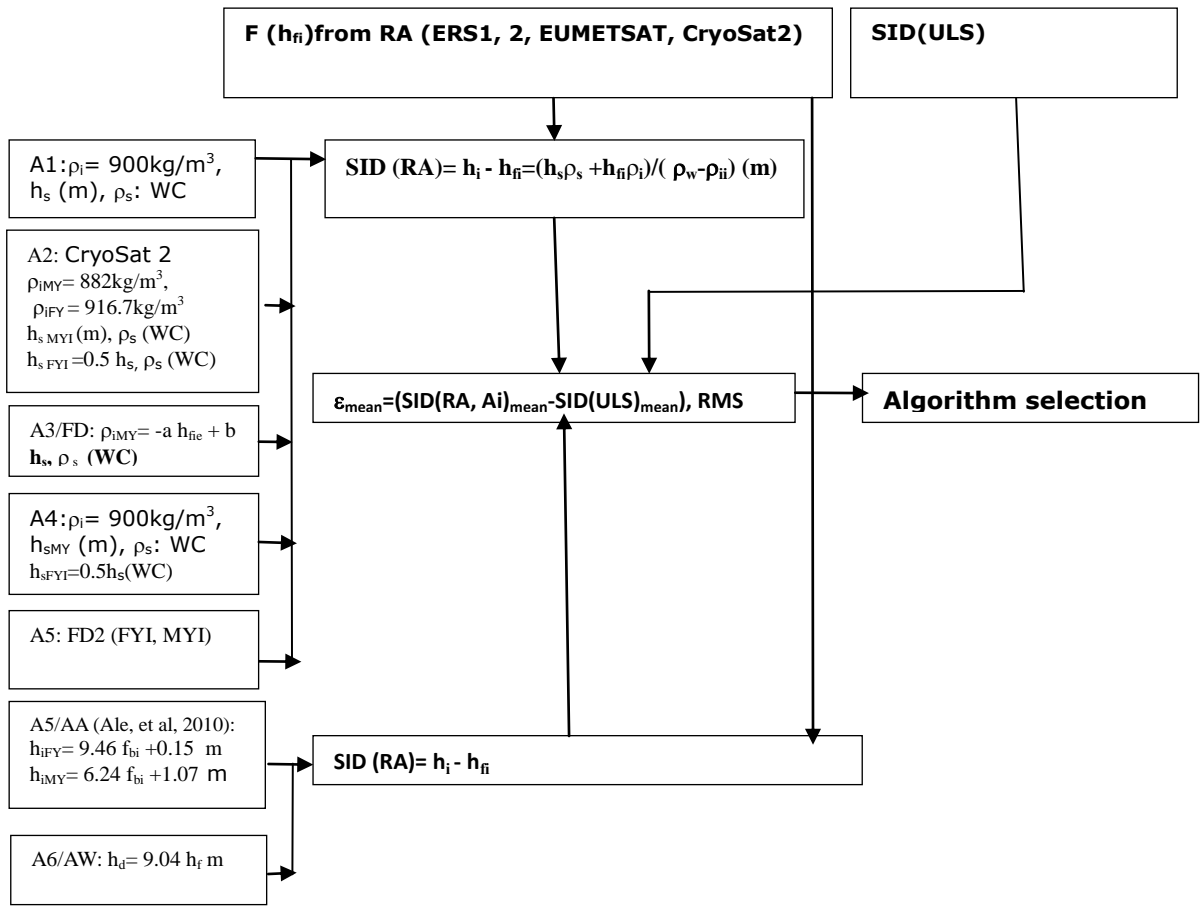

Figure 5. Algorithm selection

**Comparison of algorithms for SIT retrieval from RA in 1996**

NSIDC ULS in Beaufort Sea, from 10/1996, collocated with RA data have been used to validate algorithms (A1-A7) for SID retrieval. SID(RA) calculated from the freeboard from RA, applying algorithms A1-A7 is compared with collocated SID retrieved from ULS (SID(ULS)) Figure 6. Statistic of SID(ULS) and SID(RA), calculated by A1-A7 is given in Table 8. Algorithms A1 and A4 have the same statistical properties because it is not information for FYI on 10/1996 in the area of Beaufort Sea and A4 performs as A1 (with the same sea ice density (900kg/m$^3$), h$_s$ and $\rho_s$ from WC and $\rho_w = 1030 kg/m^3$). SID estimated by CryoSat (A2) Algorithm is again underestimated even in presence of only MYI, confirmed with negative bias and histogram analyses. The histograms (Figure 7) of residuals (SID(RA,Ai)-SID(ULS)) shows the impact of different algorithms on residuals of SID derived from RA and SID(ULS). Algorithm A(W) has maximum RMS error and a small bias. The small bias of A(W) is not confirmed with other data sets and is probably because SID(ULS) from 1996 have been used to validate the A(W) algorithm in the past. Only SID, calculated from RA, using FD (A(FD) and A(FD2)) algorithms are with minimum biases, minimum RMSE and minimum deviation (Figure 7/d).

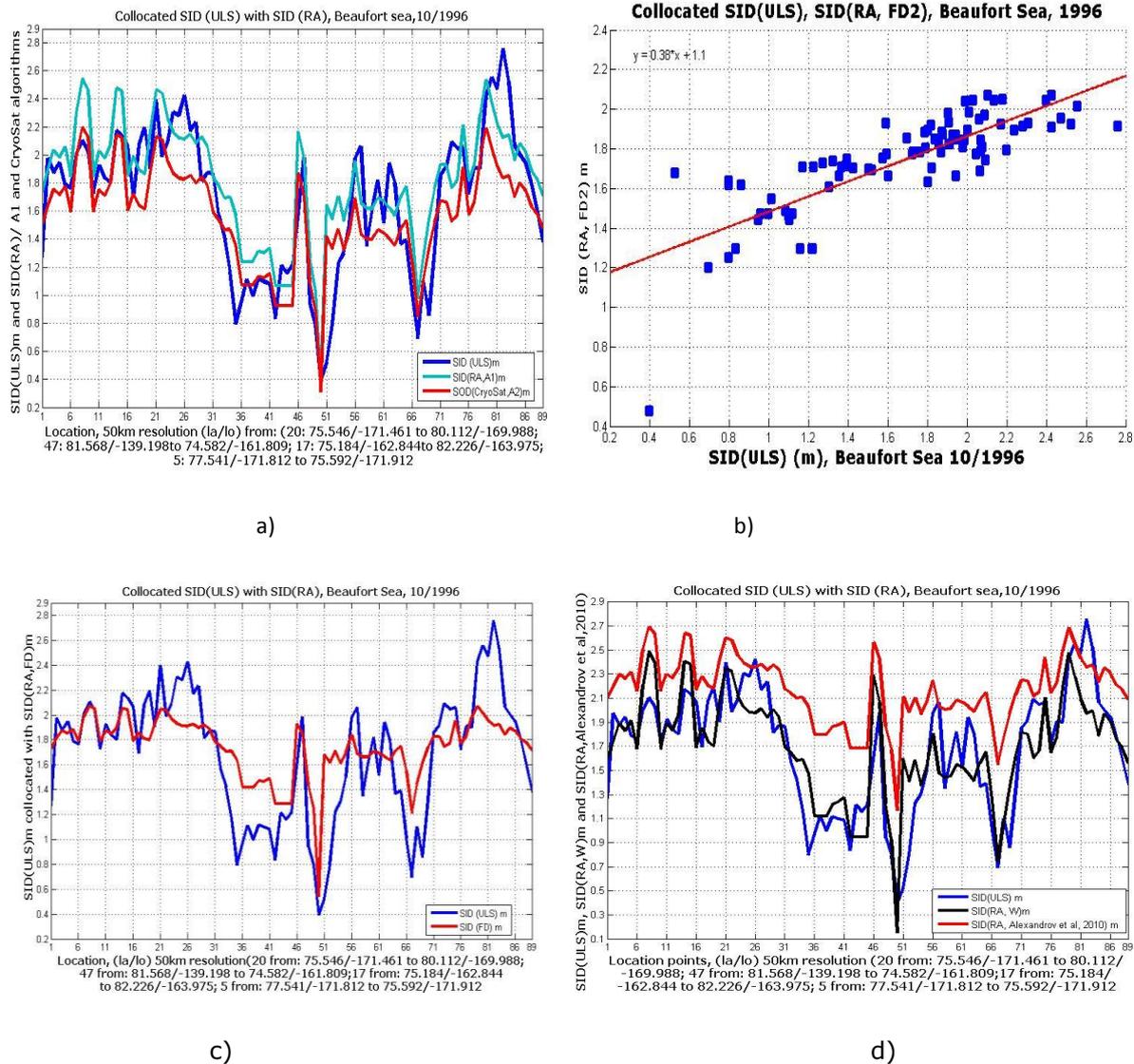

Figure 6. Collocated SID(m) from NSIDC/ULS and RA2, calculated using Algorithms A1-A7 in Beaufort Sea, 10/1996: a) SID(ULS), SID(RA, A1) and SID (RA,A2/CryoSat2); b) SID(ULS) with SID(RA,FD2); c) SID(ULS), SID(RA,FD), d) SID(ULS) with SID(RA,AA) and SID(RA,A7/AW).

The bias and RMSE are similar for the FD algorithms because it is not information for presence of FYI over this area. The mean value of SID retrieved from the both FD algorithms and corresponding confidence intervals are in the same range as that from ULS, which is not the case of all other algorithms (Figure 7/d). The linear regression (Figure 6/b) between SID(RA,FD2) and SID(ULS) is with the best RMSE in comparison with the other algorithms (Table 8) and the bias between SID retrieved from RA, applying A(FD2) algorithm and SID retrieved from ULS is minimum.

Table 8. Statistic of collocated SID(ULS) for 10/1996 and SID(RA), calculated from the RA by Algorithms A1- A7 , in Beaufort Sea over 4450km.

| (m) | d(ULS) | d(A1) | d(A2) | d(FD/ A3) | d(A4) | d(FD2/A5) | d(A6) | d(W) A7 |
|---|---|---|---|---|---|---|---|---|
| Min | 0.3936 | 0.3597 | 0.313 | 0.54 | 0.3597 | 0.4794 | 1.17 | 0.153 |
| Max | 2.76 | 2.543 | 2.19 | 2.07 | 2.543 | 2.073 | 2.69 | 2.49 |
| Mean | 1.678 | 1.813 | 1.567 | 1.741 | 1.813 | 1.74 | 2.17 | 1.6 |
| Median | 1.815 | 1.847 | 1.59 | 1.786 | 1.847 | 1.786 | 2.17 | 1.6 |
| Mode | 0.3936 | 1.067 | 0.92 | 1.287 | 1.067 | 1.29 | 1.68 | 0.94 |
| Std | 0.5141 | 0.4115 | 0.355 | 0.2437 | 0.4115 | 0.244 | 0.279 | 0.428 |
| $\varepsilon$ | 0 | 0.1351 | -0.111 | 0.063 | 0.135 | 0.062 | 0.491 | -0.08 |
| a | na | 0.664 | 0.574 | 0.382 | 0.66 | 0.3805 | 0.4314 | 0.667 |
| b | na | 0.698 | 0.604 | 1.099 | 0.698 | 1.104 | 1.44 | 0.567 |
| RMSE | na | 0.2309 | 0.199 | 0.1447 | 0.2309 | 0.14 | 0.1676 | 0.257 |

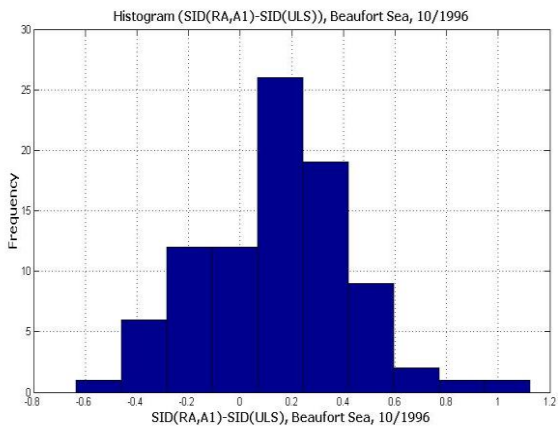
a) SID(RA,A1)-SID(ULS), $\varepsilon$= 0.135m

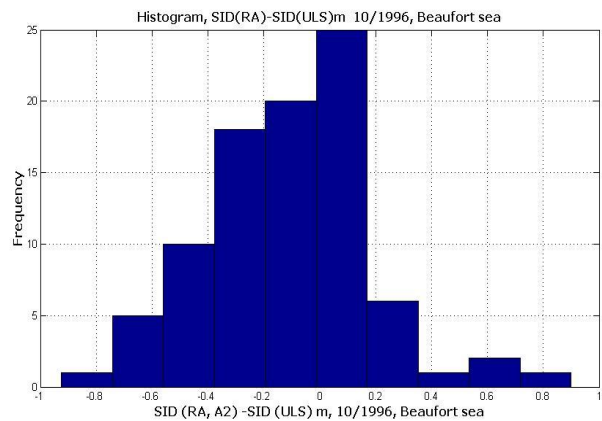
b) SID(RA,A2)-SID(ULS), $\varepsilon$=-0.111m

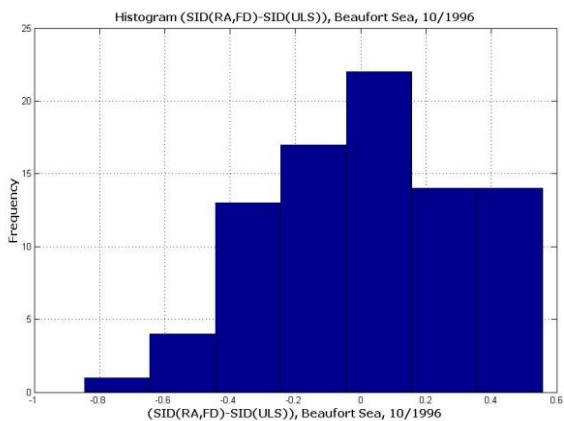
c) SID(RA,A3)-SID(ULS) m, $\varepsilon$=0.06m

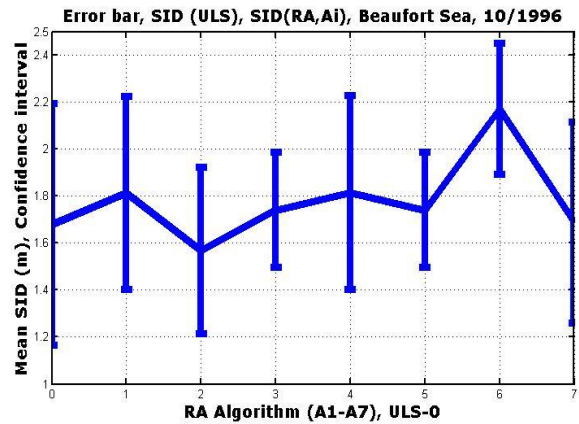
d) Error bar of the mean SID

Figure 7. Histograms of residuals (m) (a-c); d) error bar, Beaufort Sea, 10, 1996.

**Comparison of Algorithms for SIT retrieval from RA in Beaufort Gyre for 2003 to 2008**

Collocated RA2 and ULS data of SID in Beaufort Gyre for winters 2004 to 2008 have been also used for Algorithm validation and selection. The winter means of SID from ULS and RA are listed in Table 9. The difference of the mean winter SID, retrieved from RA (A1) and the mean winter SID, retrieved from ULS depends on year, season, ice type and is estimated as the bias $\varepsilon = SID_{ra} - SID_{uls}$ (Table 10). Due to very high bias ($\varepsilon = 0.71m$) (Table 10) and very low correlation between SID (ULS) and SID(RA) in 2003-2004, only data from 2004 to 3/2008 are used for algorithm selection. SID statistic of validated A1-A7 algorithms with SID derived from ULS in Beaufort Gyre (from 2004 to 2008) are listed in Table 11, histograms are shown on Figure 9, the graphs are shown on Figure 8 and the error bar is on Figure 8/b. Absence of information for the presence of FYI before 2007 or presence of snow melting before October, which prevents radar signal penetration in the snow, explains larger biases between SID(RA) and SID(ULS) in October and February. The small mean bias between SID(RA, A1) and SID(ULS) can be explained with validation of SID(RA) in the past using A1 algorithm with SID(ULS) data in Beaufort Gyre (Table 11). The histogram of the residual SID(RA, A1) and SID(ULS) is shown on Figure 9/c.

From Figure 8 and Tables 11 one can see that A2 and A4 Algorithms underestimate SID(ULS) up to -0.27m due to assumption of half snow depth over FYI and use of fixed ice density (882kg/m$^3$) over MYI (Figure 8, Table 11). The impact of assumption of $0.5h_s(WC)$ over FYI on increased negative bias between calculated SID(RA,A4) and SID (ULS) is shown in the last (35-40) months (Figure 8/a), when an information from OSI SAF for presence of FYI is collocated with the freeboard, derived from RA, and the largest biases over FYI of the retrieved SID by A2 and A4 are observed. The higher biases (Figure 9/b, d) and RMSE (Table 11) of the SID, retrieved from RA using A2(CryoSat2) and A4 Algorithms, compared with SID(ULS) over FYI confirm that the assumption for half snow depth ($0.5h_s(WC)$) over FYI is wrong, not applicable to retrieve SIT and SID from RA, using the equation for hydrostatic equilibrium and will always lead to underestimation of SIT and higher uncertainties, confirmed with sensitivity and uncertainty analyses

Table 9. Winter means SID (m) estimated in Beaufort Gyre from RA2 and ULS from 2003 to 2008

| Mean Year | RA2 | ULS | RA2 03 | ULS 03 | RA2 04 | ULS 04 | RA2 05 | ULS 05 | RA2 06 | ULS 06 | RA2 07 | ULS 07 | RA2 08 | ULS 08 |
|---|---|---|---|---|---|---|---|---|---|---|---|---|---|---|
| SID | 1.64 | 1.59 | 1.62 | 0.91 | 1.78 | 1.6 | 1.67 | 1.78 | 1.67 | 1.66 | 1.53 | 1.57 | 1.48 | 1.77 |

Table 10. Bias of the derived mean winter SID (Algorithm A1) from RA and the collocated mean winter SID, from moored ULS in Beaufort Gyre from 2003 to 2008.

| Year | 2003 | 2004 | 2005 | 2006 | 2007 | 2008 |
|---|---|---|---|---|---|---|
| SID bias(m) | 0.71 | 0.18 | -0.11 | 0.01 | -0.04 | -0.29 |

Table 11. Statistic of SID, from ULS and RA2 in Beaufort Gyre, 2004-03/2008

| Parameter | d (ULS) | d(A1) | d(A2) | d(FD), A3 | d(A4) | d(FD2),A5 | d(AA), A6 | d(W),A7 |
|---|---|---|---|---|---|---|---|---|
| Min(m) | 0.97 | 1.24 | 1.07 | 1.41 | 1.164 | 1.419 | 1.267 | 0.852 |
| Max(m) | 2.32 | 2.13 | 1.85 | 1.92 | 2.14 | 1.923 | 2.06 | 1.52 |
| Mean(m) | 1.665 | 1.60 | 1.4 | 1.656 | 1.566 | 1.658 | 1.802 | 1.176 |
| Median(m) | 1.6 | 1.59 | 1.38 | 1.655 | 1.592 | 1.652 | 1.88 | 1.174 |
| Mode(m) | 0.97 | 1.24 | 1.07 | 1.419 | 1.164 | 1.419 | 2.06 | 1.17 |
| Std(m) | 0.3535 | 0.2617 | 0.2208 | 0.146 | 0.299 | 0.1434 | 0.236 | 0.197 |
| Bias (B) | 0 | -0.07 | -0.2654 | -0.009 | -0.1 | -0.007 | 0.098 | -0.4889 |
| abg | na | 0.6215 | 0.5274 | 0.3571 | 0.6214 | 0.3517 | 0.185 | 0.2437 |
| bbg | na | 0.5741 | 0.5216 | 1.062 | 0.5395 | 1.072 | 1.455 | 0.7704 |
| RMS | na | 0.1454 | 0.1211 | 0.0753 | 0.2077 | 0.073 | 0.2454 | 0.1813 |

Only A(FD2) algorithm provide minimum bias (0.007m) between SID(RA, FD) and SID(ULS) where $\rho_i(h_{fe})$ is calculated considering the $h_{fi}$ and $h_s$ (Figure 8/d, Table 11). The sea ice density of A(FD2) is calculated as a function of effective freeboard (Equation 14-17) and the impact of ice type on $\rho_i$ is shown on Figure 8/d, where the $\rho_i(h_{fe})$, calculated over FYI is higher than that over MYI, which is confirmed by Kovacs, (1996).

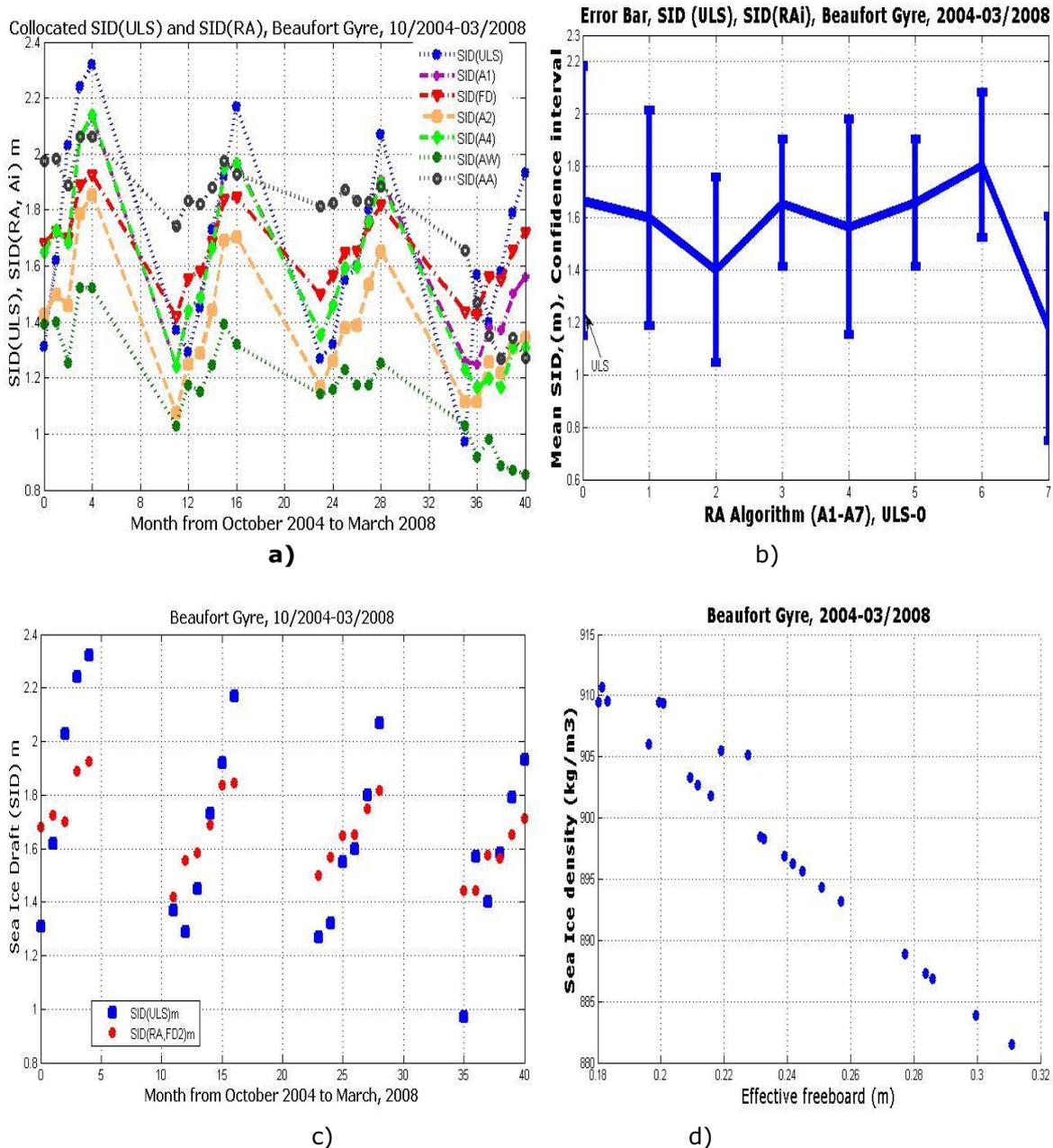

Figure 8. Validation of algorithms A1-A7 in Beaufort Gyre from October 2004 to March 2008 a) SID (RA) calculated, using algorithms A1-A4, A6-A7, collocated with SID (ULS); b) Error bar; c) Collocated SID(RA,FD2) and SID(ULS);d) Sea ice density dependence on effective freeboard.

The biases between SID(RA) (using A(FD), or A(FD2)) and SID(ULS) are improved when a collocated information from OSI-SAF for presence of FYI in the period 10/2007-2/2008 is available, which confirms the high accuracy of the FD algorithms, when correct information for ice type is provided (Figure 8/a, c). The minimum bias and RMSE between SID, retrieved from RA using A(FD2) and SID (ULS) confirms the high accuracy of A(FD2) algorithm to retrieve SIT and SID from RA over FYI and MYI.

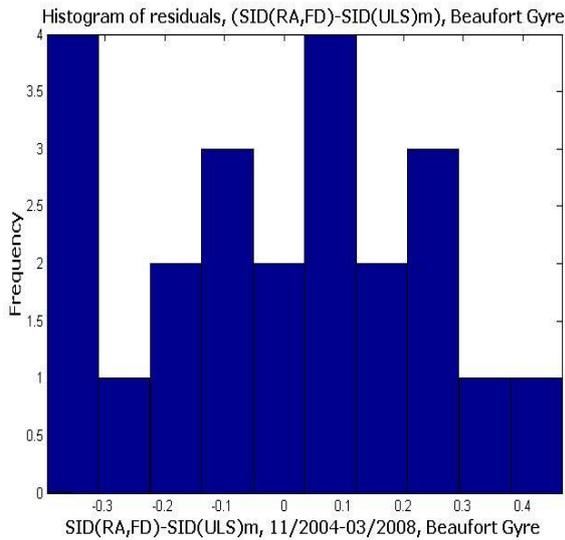
a) SID(RA, FD) − SID(ULS)

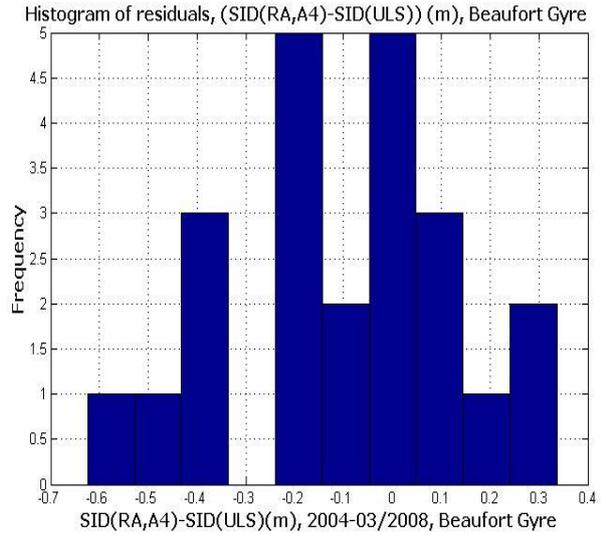
b) SID(RA, A4) - SID(ULS)

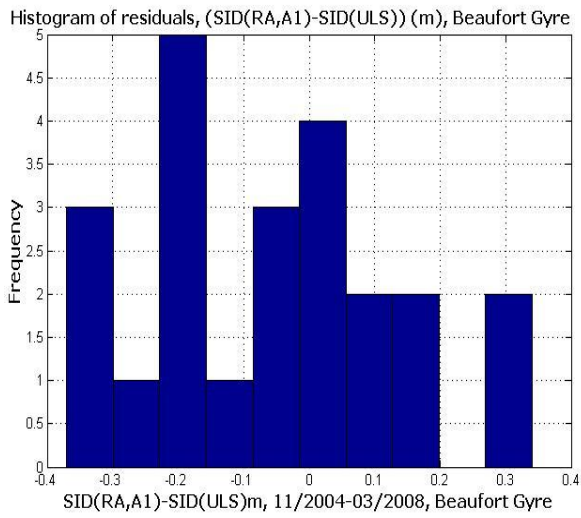
c) SID(RA, A1) − SID(ULS)

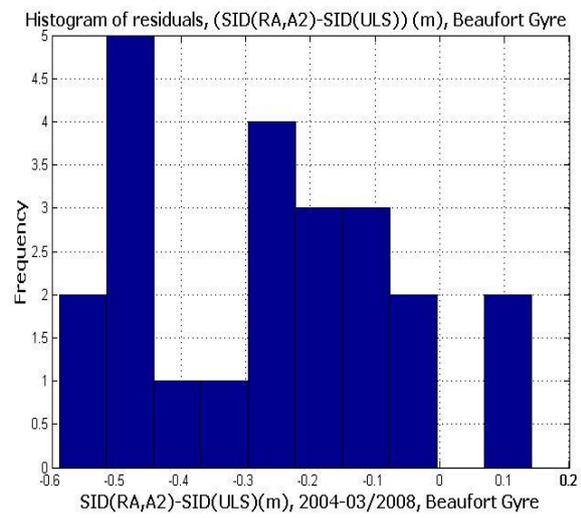
d) SID(RA, A2) - SID(ULS)

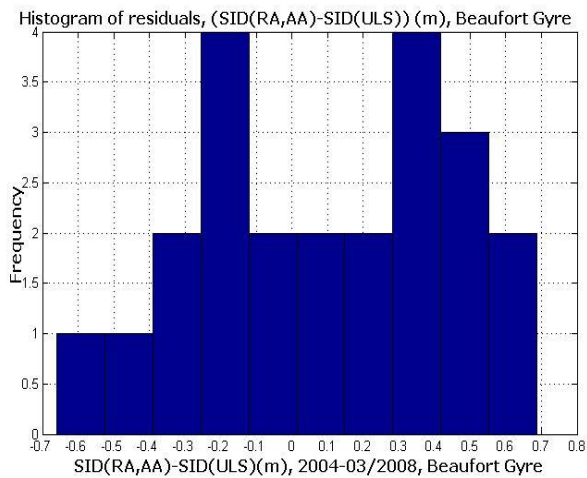
e) SID(RA, AA) − SID(ULS)

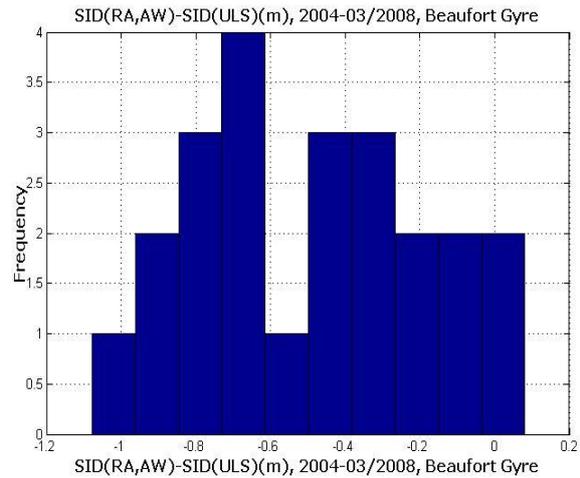
f) SID(RA, AW) - SID(ULS)

Figure 9. Histograms of residuals, Beaufort Gyre, 2004-2008

## 4. Algorithm selection and conclusions

All currently existing algorithms are validated in Section 3. Due to uncertainties, related with input variables and systematic underestimation of A2 (CryoSat-2) algorithm, 2 algorithms have been added: i) A3(FD) to test the impact of free-board dependent MYI density on accuracy of the retrieved SID and SIT from F, derived from RA; ii) A5(FD2) extended FD algorithm over FYI and MYI; iii) A4 to test the impact of assumption $h_s=0.5h_s(WC)$ over FYI on accuracy of SID in presence of FYI.

Algorithm A2 demonstrated systematic underestimation of SID compared with independent collocated ULS data within 12 years over MYI and FYI due to assumption of half snow depth ($h_s=0.5h_s(WC)$) over FYI and use of fixed sea ice density ($882 kg/m^3$) over MYI. The sensitivity analyses demonstrated that assumption of half snow depth over FYI will always lead to underestimation of SID and SIT calculated by A4 and A2 and the impact depends on $\rho_i$, $\rho_s$, $\rho_w$ and free-board used as input variables in Equation 5. It was not information for presence of FYI for area of NSIDC SID observations in 1996, Beaufort Sea and A2 (Cryosat2) algorithm calculated SIT and SID over MYI, using $h_s$ and $\rho_s$ from WC, which lead to smaller negative bias due to fixes $\rho_i$, used for MYI. Presence of FYI in Beaufort Gyre (in 2007-2008) and using the assumption of $0.5h_s(WC)$ over FYI, increase the negative bias of A2 algorithm (more than 2 times).

Algorithm A4 is the same as Algorithm 1 over MYI and this is the reason why the biases are the same for A1 and A4 using collocated ULS in 10/1996. The purpose of Algorithm A4 is to test the impact of assumption of $h_s=0.5h_s(WC)$ over FYI on accuracy of the derived SIT from RA, using fixed ice density ($900kg/m^3$). Comparison of SID, calculated by A4 (assuming $h_s=0.5h_s(WC)$ over FYI) show that even a small presence of FYI leading to assumption $h_s=0.5h_s(WC)$, inserted in the equation for hydrostatic equilibrium (Equation 5) can lead to essential underestimation of SID (Table11), which confirms that this assumption is not valid and cannot be applied to retrieve SIT and SID using the equation for hydrostatic equilibrium.

The mean biases between the SID retrieved from RA and the collocated SID from ULS, as well as the RMSE for all 7 algorithms compared in Beaufort Sea and Beaufort Gyre in the period 2006-2008 are summarised in Table 12, including also the RMSE, reported in Djepa (2013a) in Beaufort sea.

Table 12. Summary of SID (RA) (A1 – A7 Algorithms) mean biases and RMS, calculated by comparison of SID(RA) with SID(ULS) within 12 years (from 1996-2008) in Beaufort Sea and Beaufort Gyre.

| Algorithm | Location | A1 | A2 | A3/FD | A4 | A5/FD2 | A5/AA | AW |
|---|---|---|---|---|---|---|---|---|
| Bias (m) | NSIDC/10/1996 | 0.1351 | -0.111 | 0.063 | 0.135 | 0.062 | 0.492 | -0.08 |
| Bias (m) | Beaufort Gyre,2004-2008 | -0.07 | -0.2654 | -0.009 | -0.1 | -0.007 | 0.098 | -0.4889 |
| Bias(m) | Beaufort sea | -0.07 | -0.26 | 0.02 | -0.38 | -0.001 | -0.476 | -0.729 |
| RMS(m) | NSIDC/10/1996 | 0.231 | 0.199 | 0.144 | 0.231 | 0.14 | 0.168 | 0.257 |
| RMS (m) | Beaufort Gyre,2002-2008 | 0.145 | 0.1211 | 0.075 | 0.2077 | 0.073 | 0.245 | 0.1813 |
| RMS(m) | Beaufort sea | 0.143 | 0.131 | 0.123 | 0.1929 | 0.1018 | 0.1614 | 0.1606 |
| Mean RMS | 1996-2008 | 0.1730 | 0.1504 | 0.114 | 0.2105 | 0.1049 | 0.1915 | 0.1996 |

Using collocated data in Beaufort Sea and Beaufort Gyre for 12 years and criteria for minimum bias and minimum RMS error of the retrieved SID from RA, compared with SID (ULS) for all collocated data (Table 12). Algorithm A5(FD2) was selected because it demonstrated better biases in presence of FYI, it is freeboard and snow depth dependent, accounting for the impact of freeboard and snow depth on FY and MY sea ice density along the RA averaged area. Algorithm A3(FD) is performing as Algorithm 5(A(FD2) over MYI but in presence of FYI and MYI A5(FD2) calculates SIT and SID as a function of FD ice density, leading to smaller bias. Algorithms A1, A2, A4, A5 and A(W) were not selected because: i) they demonstrated unstable biases; ii) A1 depends on fixed $\rho_i$ with unknown accuracy and considering the large range of $\rho_i$ reported for MYI and FYI the biases will depend on location, month and ice type and may be different for other locations; iii) A(W) and A5 are empirical algorithms with local application and do not consider the impact of snow depth. Algorithms A2 and A4 were not selected because: i) A2 demonstrated underestimation of SID (RA) compared with SID(ULS) over MYI and FYI over 12 years, confirmed also with sensitivity analyses ; ii) A4 was not selected due to assumption of half snow depth over FYI, leading to systematic underestimation of SIT and SID calculated from RA, confirmed with validation with SID(ULS) and sensitivity analyses. The impact of the uncertainties of the variables, contributing to retrieval of SID and SIT are listed in Table 7 for all algorithms using the equation for hydrostatic equilibrium.

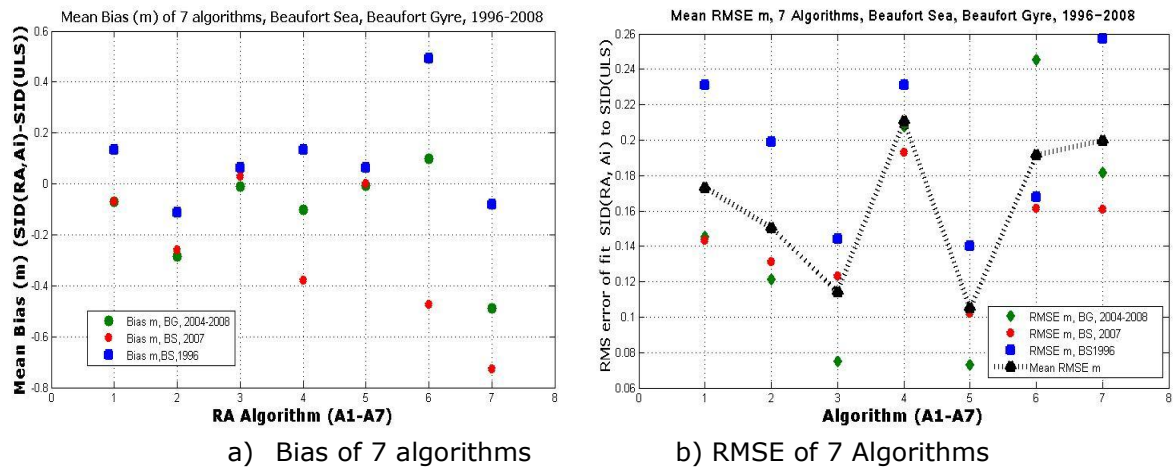

a) Bias of 7 algorithms    b) RMSE of 7 Algorithms

Figure 10. Mean Bias and RMSE of 7 Algorithms, validated to retrieve SID and SIT from RA.

The biases and RMSE for all 7 algorithms are summarised in Figure 10. One can see that Algorithm A3 (A(FD)) and A5(FD2) demonstrated stable minimum bias and RMSE over 12 years (Figure 10).

**Conclusions**

The algorithms to retrieve SIT and SID from ULS and radar altimeters are reviewed with precise analyse of uncertainties of contributing input variables and new freeboard depended algorithms to retrieve SIT from RA have been developed and validated. The impact of sea ice, snow density and depth on accuracy of the retrieved SIT from RA is analysed, applying sensitivity analyses and the uncertainties of the retrieved SIT from RA and SID from ULS are summarised. The most important results of this paper are:

 i)     The hypothesis of half snow depth over FYI has been analysed by applying sensitivity analyses and by comparison of SID, derived from RA, using different algorithm and SID retrieved from independent ULS observations over 12 years. It was confirmed (by sensitivity analyses and observations) that the assumption of half snow depth over FYI, applied for sea ice mass balance retrieval from CryoSat [Laxon et al, 2012] will always underestimate the retrieved SIT from RA when the equation for hydrostatic equilibrium is applied. The accuracy of the retrieved snow depth from OIB radar in 2009 and 2010 and the AMSR-E snow depth product are analysed and it is concluded that it is not possible to provide quantitative estimate of the snow depth retrieved from AMSR-E and the airborne radar because the accuracy of any of these algorithms is not known. Evidences of 2.3 times underestimation of snow depth, retrieved from AMSR-E algorithm [Worby, et al, 2008] and the calibration of the both instruments to show similar readings [Cavalieri et al, 2012] explain the misleading assumption for half snow depth over FYI.

ii)     The sea ice density dependence on freeboard and ice type is analysed using extended literature review, sensitivity, uncertainty analyses, observations [Ackley, 1976, Kovacs et al, 1989], freeboard dependent sea ice density algorithms to retrieve SIT from RA and comparison of SID, retrieved from RA with SID derived from ULS. The developed (over FYI and MYI) FD algorithms to retrieve SIT and SID from RA are validated with collocated SID retrieved from ULS. The algorithm selection is based on a Statistic, minimum bias and RMSE.

iii) The applicability of WC for SIT retrieval from RA, using equation for hydrostatic equilibrium has been examined using sensitivity analyses and comparison of 5 algorithms for SIT retrieval and the high accuracy (1mm bias and minimum RMSE) of FD algorithm for SIT retrieval from RA over FYI and MYI was confirmed, using snow depth and ice density from WC, which confirms the applicability of WC over FYI and MYI.

iv) An extended analysis is provided on propagated uncertainties of the retrieved SIT and SID for all 5 algorithms, considering the uncertainties of input variables.

v) v) Seven algorithms for SIT retrieval from RA have been compared, validated and a freeboard depended algorithm has been selected based on a statistic for freeboard to SIT conversion, using RA. A sea ice density, calculated as a function of freeboard over FYI and MYI, is inserted in the hydrostatic equilibrium equation to retrieve SIT and SID from F, derived from RA. SID calculated from RA, applying FD algorithm is compared with SID, derived from ULS over 12 years in Beaufort Sea and Beaufort Gyre. The FD algorithm is selected, providing SID, derived from RA with minimum RMSE and bias with SID (ULS) over 12 years in the Arctic.
The results and developed algorithms will benefit not only SIT retrieval from RA but also the SIT retrieval from laser altimeter because the both algorithms are based on hydrostatic equilibrium and depend on accuracy of the same input variables.
Considering the importance of SIT for climate, NWP models and sea ice balance in the Arctic and Antarctica the above results will benefit current and future ESA, NASA programs, climate change, numerical prediction by introducing more accurate algorithms for SIT retrieval from Radar Altimeters.